\documentclass[11pt]{article}

\usepackage{fullpage}

\usepackage{booktabs} 

\usepackage{latexsym}
\usepackage{url}
\usepackage{listings}
\usepackage{xspace}
\usepackage{color}
\usepackage{graphicx}

\usepackage{amsmath}
\usepackage{amssymb}

\newcommand{\eg}{e.g.\@\xspace}

\newcommand{\vs}{vs.\@\xspace}

\newcommand{\repetitions}{100\xspace}
\newcommand{\warmup}{5\xspace}

\newcommand{\mpibarrier}{\texttt{MPI\_\-Barrier}\xspace}
\newcommand{\mpibcast}{\texttt{MPI\_\-Bcast}\xspace}
\newcommand{\mpiscatter}{\texttt{MPI\_\-Scatter}\xspace}
\newcommand{\mpialltoall}{\texttt{MPI\_\-Alltoall}\xspace}
\newcommand{\mpiwaitall}{\texttt{MPI\_\-Waitall}\xspace}
\newcommand{\mpiwtime}{\texttt{MPI\_\-Wtime}\xspace}

\newcommand{\mpiint}{\texttt{MPI\_\-INT}\xspace}

\newcommand{\openmpiversion}{Open\,MPI 3.1.3\xspace}
\newcommand{\intelmpiversion}{Intel\,MPI 2018\xspace}
\newcommand{\mpichversion}{\texttt{mpich} 3.3\xspace}

\newcommand{\gccversion}{\texttt{gcc 8.3.0}\xspace}

\newcommand{\ceiling}[1]{\lceil #1\rceil}

\begin{document}

\title{$k$-ported \vs $k$-lane Broadcast, Scatter, and Alltoall
  Algorithms}
\author{Jesper Larsson Tr\"aff\\
TU Wien, Faculty of Informatics, Institute of Computer Engineering 191-4\\
Favoritenstrasse 16/3rd floor, 1040 Vienna, Austria}

\maketitle
\begin{abstract}
  In $k$-ported message-passing systems, a processor can
  simultaneously receive $k$ different messages from $k$ other
  processors, and send $k$ different messages to $k$ other processors
  that may or may not be different from the processors from which
  messages are received. Modern clustered systems may not have such
  capabilities. Instead, compute nodes consisting of $n$ processors
  can simultaneously send and receive $k$ messages from other nodes,
  by letting $k$ processors on the nodes concurrently send and receive
  at most one message. We pose the question of how to design good
  algorithms for this $k$-lane model, possibly by adapting algorithms
  devised for the traditional $k$-ported model.

  We discuss and compare a number of (non-optimal) $k$-lane algorithms
  for the broadcast, scatter and alltoall collective operations (as
  found in, \eg, MPI), and experimentally evaluate these on a small
  $36\times 32$-node cluster with a dual OmniPath network (corresponding to
  $k=2$). Results are preliminary.
\end{abstract}

\section{Introduction}

We pose the problem of designing good collective communication
algorithms for modern high-performance clusters. By this pompous
opener, we mean the following.  In $k$-ported message-passing systems,
a processor can simultaneously receive $k$ different messages from $k$
other processors, and send $k$ different messages to $k$ other
processors that may or may not be different from the processors from
which messages are received. In this model, many standard collective
communication operations (broadcast, allgather, reduction, scatter,
etc.) have been studied, and for many operations, algorithms designed
for one-ported (bidirectional, send-receive) communication extend
nicely and close to
optimally~\cite{Bar-Noy95,BarNoyHo99,Bruck97,SackGropp15}.  Modern
clustered systems may not have such strong capabilities per
processor-core. Instead, compute nodes consisting of $n$ processors
can simultaneously send and receive $k$ messages from other nodes, by
letting $k$ processors on the nodes concurrently send and receive at
most one message. For instance, compute nodes may have several
connections to the communication network, multiple network switches,
and in general offer more off-node network bandwidth than can be
saturated by a single processor-core on the compute node. We call such
capabilities \emph{multi-lane}, and the \emph{$k$-lane model} allows
$k$ simultaneous communication operations per $n$-processor-core
compute node, possibly concurrently with communication inside the
compute node via shared memory.

We examine two approaches to the design of good algorithms for
$k$-lane systems. One is based on splitting the problem data across
$n$ process-cores and apply standard one-ported algorithms
concurrently. This approach has been pursued in our previous
work~\cite{Traff19:lanecorr,Traff20:mpidecomp}. The other approach
adapts $k$-ported algorithm to the $k$-lane model by additional
communication on the compute nodes to distribute problem data to $k$
node local processor-cores. We make the discussion concrete by
describing initial algorithms for the broadcast, scatter and alltoall
collective operations as found in the MPI operations \mpibcast,
\mpiscatter and \mpialltoall~\cite{MPI-3.1}. We have implemented these
algorithms and compare against standard $k$-ported
algorithms~\cite{ChanHeimlichPurkayasthavandeGeijn07} and the
corresponding native MPI library implementations on a small, $36\times
32$-core cluster with dual-switch communication network and different
MPI libraries.

Our experimental results so far are inconclusive. Efficiently
exploiting $k$-lane communication seems to require also strong
shared-memory support and efficient communication on the compute nodes
as well as concurrent on- and off-node communication. Also, both
algorithms and the implementations in MPI can be
improved. Nevertheless, we think the $k$-lane model deserves further
investigation towards better balancing on- and off-node
communication capabilities for improved performance of collective
communication operations.

\section{Algorithms}
\label{sec:algorithms}

We assume a high-performance computing system with $p$ processor-cores
located on $N$ compute nodes, each with $n$ processor-cores (on one or
more sockets, which will play no role at the moment), such that
$p=Nn$.  Each processor has a unique \emph{rank} $i$, and ranks are
consecutive, $0\leq i <p$.

We consider algorithms for broadcast (mostly for small data sizes),
scatter and alltoall. In the \emph{broadcast operation}, $c$ data
elements from a designated root processor $r$ has to be disseminated
to all $p$ processors. In the \emph{scatter operation}, the designated
root processor has an individual block of data $b_i$ to be
communicated to each of the (other) processors with each block being
of size $c/p$ data elements. The \emph{gather operation} is the dual
of the scatter operation, and not treated further here. In the
\emph{personalized alltoall operation} each processor has an
individual block of data of size $c/p$ elements to be communicated to
each of the (other) processors; equivalently, each processor has to
receive a block of $c/p$ elements from each of the (other) processors.
In MPI, the regular broadcast, scatter and alltoall operations are
performed by the collective operations \mpibcast, \mpiscatter, and
\mpialltoall called on a \emph{communicator} with $p$ MPI
processes~\cite{MPI-3.1}.

\subsection{Standard $k$-ported algorithms}
\label{sec:k-ported-algorithms}

A straightforward divide and conquer approach leads to $k$-ported
algorithms for both broadcast and scatter operations, see,
\eg,~\cite{ChanHeimlichPurkayasthavandeGeijn07}.

Let $p$ be the number of processors, and let $k, k\geq 1$ be the
number of communication ports per processor, which means that a
processor can be involved in $k$ send and receive operations at a
time. Each processor $i$ will maintain a range of processors $[s,e-1]$
to which it belongs, $0\leq i<e$, and a root processor $r$, $s\leq
r<e$ also in the range. Initially, all processors start out with the
same range $[0,p-1]$ and $r$ being the given root of the broadcast or
scatter operation. In the divide step, each processor divides its
range $[s,e-1]$ into $k+1$ roughly equal-sized subranges
$[s_i,s_{i+1}-1]$, $i=0,1,k$, $s_i<s_{i+1}$, differing in size by at
most one. The root processor $r$ which belongs to one of the
subranges, say $[s_j,s_{j+1}-1]$ sends data to a new, local root $r_i$
in each of the other subranges. For instance, $r_i$ could be chosen as
$s_i$ (except for the range $i=j$). Each processor continues to the
next divide step with the range $[s_i,s_{i+1}-1]$ and local root
$r_i\in [s_i,s_{i+1}-1]$ to which it belongs.  A processor terminates
when its range consists only of the processor itself, at which point
at the latest it will have become a local root and received data from
some other processor.

For the broadcast algorithm, a (local) root processor always sends all
(the same) $c$ data elements to the new subroots. For the scatter
algorithm, the root sends the data blocks $[b_{s_i}b_{s_i+1}\ldots
  b_{s_{i+1}-1}]$ to the local root $r_i$ of subrange $[s_i,s_{i-1}-1]$.

Obviously, these algorithms take $\ceiling{\log_{k+1}p}$ communication
rounds.  Broadcast in each round sends $c$ data elements, while
scatter sends $c$ elements in total over all communication rounds.
Thus, this $k$-ported broadcast algorithm is only good for small
element counts $c$, since broadcast can be done in $O(\log p+c)$
rounds and communication time by other, different
algorithms~\cite{Jia09,Traff08:optibcast}. The scatter algorithm sends
the total number of data elements $c$ only once from the root
processor, and is thus round and message size optimal.

For the alltoall operation, the straightforward algorithm does
$\ceiling{p/k}$ communication rounds, in each of which each processor
sends its $k$ blocks of data to the $k$ ``next'' processors and
receives its blocks from the ``previous'' $k$-processors. This
algorithm is message size optimal in the sense that each block is sent
and received exactly once. Message-combining algorithms and
implementations~\cite{Bruck97,Traff14:bruck} can reduce the number of
communication rounds (to $\ceiling{\log_{k+1} p}$) at the cost of
sending and receiving more than $c$ data elements.

\subsection{$k$-lane algorithms: Approach problem splitting}

A straight-forward approach to exploit multi-lane communication
capabilities is to split the collective communication problem to be
solved on $c$ data elements into $n$ (similar) independent subproblems
to be solved on $c/n$ data elements. The splitting is performed
concurrently on one, more, or all of the $N$ compute
nodes. With $k$-lane communication capabilities, the $n$ independent
subproblems on $c/n$ data elements can ideally be solved with a
speed-up of a factor of $k$. The overall speed-up is determined by
this, and the amount of work and time to be spent in the splitting of
the $c$ element problem on the compute nodes.

Concretely, for the broadcast problem, the compute node at which the
designated broadcast root $r$ is residing splits the $c$ data elements
over the $n$ processors on the node into blocks of $n/c$ elements (one
block per processor) by a scatter operation on the node. After that,
$n$ concurrent broadcast operations each on $N$ processors on
different nodes are performed. To complete, on all nodes, the blocks of
$c/n$ elements need to be collected together to a block of $c$
elements which must be distributed to all processors on the node. This
is a collective allgather operation. Assuming optimal algorithms for
all component operations (scatter on root node, broadcast between
nodes, allgather on compute nodes), this algorithm has an overhead
penalty by the final allgather operations (extra communication rounds,
extra data communication on the compute nodes). A scatter algorithm,
in contrast, following the same idea, is optimal in terms of
communication rounds and data communicated. First, a scatter operation
is done on the node on which the designated root processor $r$ resides
into $n$ scatter problems of $c/n$ data elements.  These $n$ scatter
problems are solved concurrently over the $N$ nodes, leaving each
processor with a data block of $c/(nN)=c/p$ data elements. Assuming
standard (optimal, one-ported) logarithmic round scatter algorithms for the
component problems, the total number of communication rounds is
$\ceiling{\log n}+\ceiling{\log N}\leq \ceiling{\log p}+1$, which is
at most one round off from optimal. The amount of data leaving the
root node is $c-c/N$.

The alltoall operation can also be implemented with this approach by
combining blocks to the same node into larger blocks of $n c/p = c/N$
data elements. This is done by concurrent alltoall operations on all
compute nodes, followed by $n$ independent alltoall operations over
the $N$ nodes. Thus, the complete data is communicated twice.

These algorithms are called \emph{full-lane} algorithms
in~\cite{Traff19:lanecorr,Traff20:mpidecomp} where they are described
in full detail with experimental results. The aim is a speed-up of a
factor of $k$ on systems utilizing $k$ physical lanes instead of just
one physical lane. Whether this is possible depends on the speed-up of
the part of the algorithms that is performed on the compute nodes.

\subsection{$k$-lane algorithms: Approach $k$-ported algorithm reuse}

Another approach to designing good algorithms for multi-lane systems is
to adapt and reuse algorithms designed for multi-ported systems. The
idea is that send and receive operations to and from $k$ other
processors by a single processor is split among $k$ processors on a
compute node that together perform the communication operations of a
single, $k$-ported processor. This in most cases entails (additional)
communication on the nodes to split and assemble larger data blocks.

For (non-pipelined) tree algorithms as often used for broadcast (for
small problem sizes) and scatter, we propose a construction that makes
use of standard $k$-ported algorithms. We exemplify with
broadcast. Let $r$ be the root processor.  We let $r$ send data first
to $k-1$ non-root processors on the same node, for instance by an
efficient broadcast algorithm for the communication capabilities of
the node. Now, $k$ processors on the root node have the data, and can
in the following step concurrently send data to $k$ other nodes. This
is done using the pattern of the $k$-ported algorithm, with each
processor taking the role of one port, and sending data to, say, the
first processor of a new node. The first time a processor on a
non-root node receives data, it first broadcasts the data to $k$ other
processors on the node, which then continue as in the $k$-ported
algorithm. The number of steps of the $k$-lane algorithm derived in
this way from a $k$-ported algorithm is at most twice that of the
$k$-ported algorithm, if we count the node local broadcast over $k$
processors as one step. Finally, to complete the broadcast, each of
the $k$ local root processors on each of the compute nodes, broadcast
their data block to $n/k$ (disjoint) processors on the node. The
$k$-lane scatter algorithm can be adopted in the same way. A receiving
processor on a node scatters to $k-1$ processors which then
concurrently do the $k$ send operations of a single processor in the
$k$-ported algorithm to processors on other nodes.

For alltoall, a $k$-lane algorithm could look as follows: In $N-1$
communication rounds, all $n$ processors on a compute node
concurrently send their $n$ data blocks to processors on the ``next''
node and receive their $n$ data blocks from processors on the
``previous'' node. In a final round, all processors on each compute
node exchanges their blocks with the other processors on the node
(on-node, local alltoall). The communication in a round is done in $n$
steps in such a way that in each step the $n$ processors on a node
send and receive from different processors. This algorithm will in
each of the $N-1$ rounds utilize the full off-node communication
bandwidth possible with $k$ lanes.

\subsection{A $k$-lane model}

What can be expected from algorithms designed as outlined above? For
the broadcast and scatter problems, both approaches entail
communication on the compute nodes, as well as $k$- or $n$-way
concurrent communication between nodes. For $n$-way communication
between nodes, we can possibly and realistically assume,
see~\cite{Traff19:lanecorr,Traff20:mpidecomp}, that bandwidth is
equally shared among the processors, that is a $k/n$-way increase
(with $k$) per processor. In order for a $k$-lane algorithm to exhibit
a $k$-fold speed-up, the part of the problem solved on the nodes must
also be sped up by a factor of $k$. Whether this is possible also
depends on the communication performance and the communication
capabilities on the compute nodes.

Communication between processors on the same node is typically via
shared memory. How much communication can the shared memory sustain?
Can all processors communicate at the same time achieving the same
(memory) bandwidth as when only one processor is reading and writing
from and to the memory? Is compute node communication like a fully
connected network with some number $k'$ of simultaneous communication
ports? How large is $k'$ compared to the number of lanes $k$? Can
communication on the node and off the node be done simultaneously? Is
the bandwidth (and latency) for off-node communication much different
from the shared-memory bandwidth for node local communication (often,
they are in the same ballpark)? Experimental work is needed to
determine reasonable assumptions; the results in
Section~\ref{sec:results} may provide some useful information.

Another way to pose the question is: What would be required from the
node local communication, bandwidth and capability wise, in order to
make it possible to design algorithms with a provable speed-up of $k$?

\section{Implementations}

We have implemented the $k$-ported algorithms, the full-lane
algorithms, and the adapted $k$-lane algorithms described in
Section~\ref{sec:algorithms} in MPI. The code is available from the
author.

To perform $k$-ported communication, we post $k$ non-blocking MPI send
and/or receive operations, followed by an \mpiwaitall, and assume that
the MPI library will be able to use the available ports for the
processor effectively. For the adapted $k$-ported algorithms, the
compute node local broadcast and scatter operations are done by the
available \mpibcast and \mpiscatter operations. The adapted broadcast
implementation does a full \mpibcast to all processors on the node
when a block of data is received by the local root, and not a $k$-way
broadcast followed by $k$ $n/k$-way broadcasts. This could,
theoretically (depending on model and implementation), slow down the
implementation. 

\section{Experimental results}
\label{sec:results}

\begin{table}
  \caption{Systems (hardware and software) used for the experimental
    evaluation.}
  \label{tab:systems}
  \begin{center}
    \begin{tabular}{crrrccc}
      Name & $n$ & $N$ & $p$ & Processor & Interconnect & MPI library \\
      \toprule
      Hydra & 32 & 36 & 1152 & Intel Xeon Gold 6130, 2.1
      GHz & Dual OmniPath & \openmpiversion \\
      & & & & & & \intelmpiversion \\
      & & & & & & \mpichversion \\
      & & & & & & with \gccversion \\
      \bottomrule
    \end{tabular}
  \end{center}
\end{table}

We have conducted a number of experiments, comparing the algorithms
against each other and against the native MPI library implementations
of \mpibcast, \mpiscatter and \mpialltoall on current, standard MPI
libraries. The results listed here have been collected on our own
small $36\times 32$-core cluster described in
Table~\ref{tab:systems}. This systems has powerful, multi-lane,
off-node communication capabilities: Each of the two sockets on the
shared-memory compute nodes has a connection to an own OmniPath
network (dual-network). Thus, at least two processors (or MPI
processes), possibly more, can at the same time communicate with
processors (or processes) on other nodes at the same cost as when only
one processor (or process) is communicating. For the MPI
implementations, we assume that processes are placed alternatingly on
the two sockets, such that, \eg, MPI processes with rank 0 and rank 1
are place on different sockets, and each close to the network
interface on the corresponding socket.

All experiments have been conducted on the full system with $N=36$
(nodes) and $n=32$ (processor-cores per node). For the $k$-ported and
the adapted $k$-lane implementations, we have used $k=1,2,3,4,5,6$ for
the number of virtual lanes. We have used three different MPI
libraries, namely \openmpiversion, \intelmpiversion, and \mpichversion.

\subsection{Compute node \vs network performance, $N=1,n=32$ \vs $N=32,n=1$}

An initial experiment illustrates the differences in capabilities
between communication on the compute nodes and between compute
nodes. We perform an alltoall operation on $p=32$ MPI processes,
either all on a single compute node ($N=1, n=32$), or on $N=32$
compute nodes with one MPI process on each ($N=32,n=1$). We use either
the $k$-ported alltoall implementation described in
Section~\ref{sec:k-ported-algorithms} or the native MPI library
\mpialltoall implementation. For the $k$-ported experiment, we choose
$k=32$ which just means that each MPI process posts $k=32$
non-blocking send and receive operations over $\ceiling{p/k}$
communication rounds. As the broadcast experiments will show
(Section~\ref{sec:bcastexperiments}), this performs better than
using blocking send-receive operations.

The count $c$ is the number of data elements per process. Data
elements are here \mpiint.  For the running times (in $\mu$-seconds,
measured with \mpibarrier and \mpiwtime), we report both average and
minimum time of the slowest process over \repetitions repetitions with
\warmup initial, not measured warm-up repetitions.

The results with \openmpiversion in are shown in
Table~\ref{tab:alltoall.p32.openmpi} and
Table~\ref{tab:alltoall.p32.openmpi.B}.

\begin{table}[h]
  \caption{Results for $k$-ported alltoall implementations on the
    ``Hydra'' system.  The MPI library used is \openmpiversion.}
    \label{tab:alltoall.p32.openmpi}
    \begin{small}
    \begin{center}
    \begin{tabular}{rrrrrrr}
      \toprule
$k$ & $n$ & $N$ & $p$ & $c$ & avg ($\mu s$) & min ($\mu s$) \\
      \midrule
      \multicolumn{7}{c}{$k$-ported alltoall $N=32, k=32$} \\
32 & 1 & 32 & 32 & 1 & 20.14 & 17.79 \\
32 & 1 & 32 & 32 & 2 & 19.40 & 17.84 \\
32 & 1 & 32 & 32 & 4 & 26.41 & 24.86 \\
32 & 1 & 32 & 32 & 19 & 26.81 & 25.37 \\
32 & 1 & 32 & 32 & 32 & 27.21 & 25.69 \\
32 & 1 & 32 & 32 & 188 & 29.56 & 27.45 \\
32 & 1 & 32 & 32 & 313 & 32.35 & 29.74 \\
32 & 1 & 32 & 32 & 1875 & 72.78 & 64.97 \\
32 & 1 & 32 & 32 & 3125 & 108.60 & 98.73 \\
32 & 1 & 32 & 32 & 18750 & 307.48 & 293.51 \\
32 & 1 & 32 & 32 & 31250 & 448.03 & 416.71 \\
\midrule
\multicolumn{7}{c}{$k$-ported alltoall $N=1, k=32$} \\
32 & 32 & 1 & 32 & 1 & 17.85 & 15.52 \\
32 & 32 & 1 & 32 & 2 & 17.48 & 15.74 \\
32 & 32 & 1 & 32 & 4 & 17.85 & 16.52 \\
32 & 32 & 1 & 32 & 19 & 22.14 & 19.78 \\
32 & 32 & 1 & 32 & 32 & 23.28 & 20.37 \\
32 & 32 & 1 & 32 & 188 & 60.54 & 50.24 \\
32 & 32 & 1 & 32 & 313 & 63.43 & 58.75 \\
32 & 32 & 1 & 32 & 1875 & 995.89 & 971.63 \\
32 & 32 & 1 & 32 & 3125 & 1389.12 & 1364.08 \\
32 & 32 & 1 & 32 & 18750 & 4744.03 & 4690.23 \\
32 & 32 & 1 & 32 & 31250 & 4618.21 & 4526.08 \\
    \bottomrule
    \end{tabular}
    \end{center}
    \end{small}
\end{table}

\begin{table}
  \caption{Results for \mpialltoall on the
    ``Hydra'' system.  The MPI library used is \openmpiversion.}
    \label{tab:alltoall.p32.openmpi.B}
    \begin{small}
    \begin{center}
    \begin{tabular}{rrrrrrr}
      \toprule
$k$ & $n$ & $N$ & $p$ & $c$ & avg ($\mu s$) & min ($\mu s$) \\
      \midrule
      \multicolumn{7}{c}{\mpialltoall $N=32$} \\
32 & 1 & 32 & 32 & 1 & 16.24 & 14.04 \\
32 & 1 & 32 & 32 & 2 & 16.35 & 14.19 \\
32 & 1 & 32 & 32 & 4 & 16.00 & 14.23 \\
32 & 1 & 32 & 32 & 19 & 18.42 & 17.02 \\
32 & 1 & 32 & 32 & 32 & 20.20 & 18.65 \\
32 & 1 & 32 & 32 & 188 & 30.12 & 28.44 \\
32 & 1 & 32 & 32 & 313 & 32.03 & 30.61 \\
32 & 1 & 32 & 32 & 1875 & 128.12 & 111.70 \\
32 & 1 & 32 & 32 & 3125 & 218.87 & 168.62 \\
32 & 1 & 32 & 32 & 18750 & 1975.13 & 1697.04 \\
32 & 1 & 32 & 32 & 31250 & 2087.67 & 1957.50 \\
\midrule
    \multicolumn{7}{c}{\mpialltoall $N=1$} \\
32 & 32 & 1 & 32 & 1 & 12.45 & 11.62 \\
32 & 32 & 1 & 32 & 2 & 12.93 & 11.87 \\
32 & 32 & 1 & 32 & 4 & 13.32 & 12.22 \\
32 & 32 & 1 & 32 & 19 & 16.03 & 15.05 \\
32 & 32 & 1 & 32 & 32 & 18.62 & 17.73 \\
32 & 32 & 1 & 32 & 188 & 50.53 & 47.71 \\
32 & 32 & 1 & 32 & 313 & 59.55 & 57.39 \\
32 & 32 & 1 & 32 & 1875 & 957.33 & 942.80 \\
32 & 32 & 1 & 32 & 3125 & 1332.91 & 1298.90 \\
32 & 32 & 1 & 32 & 18750 & 4541.63 & 4483.26 \\
32 & 32 & 1 & 32 & 31250 & 4400.47 & 4309.64 \\
    \bottomrule
    \end{tabular}
    \end{center}
    \end{small}
\end{table}

The results with \intelmpiversion in are shown
Table~\ref{tab:alltoall.p32.intel} and
Table~\ref{tab:alltoall.p32.intel.B}.

\begin{table}
  \caption{Results for the $k$-ported alltoall implementation on the
    ``Hydra'' system.  The MPI library used is \intelmpiversion.}
    \label{tab:alltoall.p32.intel}
    \begin{small}
    \begin{center}
    \begin{tabular}{rrrrrrr}
      \toprule
$k$ & $n$ & $N$ & $p$ & $c$ & avg ($\mu s$) & min ($\mu s$) \\
      \midrule
      \multicolumn{7}{c}{$k$-ported alltoall $N=32, k=32$} \\
32 & 1 & 32 & 32 & 1 & 22.14 & 20.03 \\
32 & 1 & 32 & 32 & 2 & 22.26 & 20.03 \\
32 & 1 & 32 & 32 & 4 & 30.08 & 28.85 \\
32 & 1 & 32 & 32 & 19 & 33.43 & 30.99 \\
32 & 1 & 32 & 32 & 32 & 31.66 & 30.04 \\
32 & 1 & 32 & 32 & 188 & 34.60 & 32.90 \\
32 & 1 & 32 & 32 & 313 & 37.23 & 34.81 \\
32 & 1 & 32 & 32 & 1875 & 73.81 & 67.95 \\
32 & 1 & 32 & 32 & 3125 & 115.36 & 105.86 \\
32 & 1 & 32 & 32 & 18750 & 306.92 & 293.97 \\
32 & 1 & 32 & 32 & 31250 & 443.57 & 421.05 \\
\midrule
    \multicolumn{7}{c}{$k$-ported alltoall $N=1, k=32$} \\
32 & 32 & 1 & 32 & 1 & 28.15 & 25.99 \\
32 & 32 & 1 & 32 & 2 & 27.06 & 25.03 \\
32 & 32 & 1 & 32 & 4 & 28.15 & 25.99 \\
32 & 32 & 1 & 32 & 19 & 32.88 & 30.99 \\
32 & 32 & 1 & 32 & 32 & 34.74 & 32.90 \\
32 & 32 & 1 & 32 & 188 & 39.97 & 37.91 \\
32 & 32 & 1 & 32 & 313 & 51.54 & 49.83 \\
32 & 32 & 1 & 32 & 1875 & 172.97 & 168.09 \\
32 & 32 & 1 & 32 & 3125 & 271.59 & 265.84 \\
32 & 32 & 1 & 32 & 18750 & 2863.88 & 2761.84 \\
32 & 32 & 1 & 32 & 31250 & 2797.51 & 2711.06 \\
    \bottomrule
    \end{tabular}
    \end{center}
    \end{small}
\end{table}

\begin{table}
  \caption{Results for \mpialltoall on the
    ``Hydra'' system.  The MPI library used is \intelmpiversion.}
    \label{tab:alltoall.p32.intel.B}
    \begin{small}
    \begin{center}
    \begin{tabular}{rrrrrrr}
      \toprule
$k$ & $n$ & $N$ & $p$ & $c$ & avg ($\mu s$) & min ($\mu s$) \\
      \midrule
      \multicolumn{7}{c}{\mpialltoall $N=32$} \\
32 & 1 & 32 & 32 & 1 & 17.68 & 15.02 \\
32 & 1 & 32 & 32 & 2 & 20.07 & 17.88 \\
32 & 1 & 32 & 32 & 4 & 20.23 & 18.12 \\
32 & 1 & 32 & 32 & 19 & 32.37 & 30.99 \\
32 & 1 & 32 & 32 & 32 & 32.39 & 30.99 \\
32 & 1 & 32 & 32 & 188 & 34.66 & 32.90 \\
32 & 1 & 32 & 32 & 313 & 37.24 & 35.05 \\
32 & 1 & 32 & 32 & 1875 & 74.38 & 70.81 \\
32 & 1 & 32 & 32 & 3125 & 108.28 & 105.86 \\
32 & 1 & 32 & 32 & 18750 & 1949.99 & 1838.92 \\
32 & 1 & 32 & 32 & 31250 & 2135.67 & 2060.89 \\
\midrule
    \multicolumn{7}{c}{\mpialltoall $N=1$} \\
32 & 32 & 1 & 32 & 1 & 11.20 & 10.97 \\
32 & 32 & 1 & 32 & 2 & 11.31 & 10.97 \\
32 & 32 & 1 & 32 & 4 & 11.68 & 10.97 \\
32 & 32 & 1 & 32 & 19 & 15.16 & 14.07 \\
32 & 32 & 1 & 32 & 32 & 17.60 & 16.93 \\
32 & 32 & 1 & 32 & 188 & 42.87 & 40.05 \\
32 & 32 & 1 & 32 & 313 & 61.50 & 56.03 \\
32 & 32 & 1 & 32 & 1875 & 192.50 & 182.87 \\
32 & 32 & 1 & 32 & 3125 & 286.89 & 278.00 \\
32 & 32 & 1 & 32 & 18750 & 2576.05 & 2498.87 \\
32 & 32 & 1 & 32 & 31250 & 3043.99 & 2959.01 \\
    \bottomrule
    \end{tabular}
    \end{center}
    \end{small}
\end{table}

The results with \mpichversion are shown in
Table~\ref{tab:alltoall.p32.mpich} and
Table~\ref{tab:alltoall.p32.mpich.B}.

\begin{table}
  \caption{Results for $k$-ported alltoall implementations on the
    ``Hydra'' system.  The MPI library used is \mpichversion.}
    \label{tab:alltoall.p32.mpich}
    \begin{small}
    \begin{center}
    \begin{tabular}{rrrrrrr}
      \toprule
$k$ & $n$ & $N$ & $p$ & $c$ & avg ($\mu s$) & min ($\mu s$) \\
      \midrule
      \multicolumn{7}{c}{$k$-ported alltoall $N=32, k=32$} \\
32 & 1 & 32 & 32 & 1 & 22.50 & 20.50 \\
32 & 1 & 32 & 32 & 2 & 22.34 & 20.74 \\
32 & 1 & 32 & 32 & 4 & 29.31 & 28.13 \\
32 & 1 & 32 & 32 & 19 & 33.66 & 32.19 \\
32 & 1 & 32 & 32 & 32 & 34.03 & 32.42 \\
32 & 1 & 32 & 32 & 188 & 36.26 & 34.57 \\
32 & 1 & 32 & 32 & 313 & 38.54 & 36.95 \\
32 & 1 & 32 & 32 & 1875 & 79.35 & 72.72 \\
32 & 1 & 32 & 32 & 3125 & 119.47 & 106.10 \\
32 & 1 & 32 & 32 & 18750 & 322.25 & 304.22 \\
32 & 1 & 32 & 32 & 31250 & 457.81 & 424.15 \\
\midrule
      \multicolumn{7}{c}{$k$-ported alltoall $N=1,k=32$} \\
32 & 32 & 1 & 32 & 1 & 52.31 & 50.31 \\
32 & 32 & 1 & 32 & 2 & 50.57 & 47.92 \\
32 & 32 & 1 & 32 & 4 & 52.00 & 49.59 \\
32 & 32 & 1 & 32 & 19 & 53.29 & 50.54 \\
32 & 32 & 1 & 32 & 32 & 53.46 & 50.78 \\
32 & 32 & 1 & 32 & 188 & 62.93 & 59.60 \\
32 & 32 & 1 & 32 & 313 & 74.74 & 71.76 \\
32 & 32 & 1 & 32 & 1875 & 183.63 & 156.40 \\
32 & 32 & 1 & 32 & 3125 & 241.62 & 214.58 \\
32 & 32 & 1 & 32 & 18750 & 1717.58 & 1653.43 \\
32 & 32 & 1 & 32 & 31250 & 2782.33 & 2729.89 \\
    \bottomrule
    \end{tabular}
    \end{center}
    \end{small}
\end{table}

\begin{table}
  \caption{Results for \mpialltoall on the
    ``Hydra'' system.  The MPI library used is \mpichversion.}
    \label{tab:alltoall.p32.mpich.B}
    \begin{small}
    \begin{center}
    \begin{tabular}{rrrrrrr}
      \toprule
$k$ & $n$ & $N$ & $p$ & $c$ & avg ($\mu s$) & min ($\mu s$) \\
      \midrule
      \multicolumn{7}{c}{\mpialltoall $N=32$} \\
32 & 1 & 32 & 32 & 1 & 19.78 & 18.12 \\
32 & 1 & 32 & 32 & 2 & 23.77 & 21.70 \\
32 & 1 & 32 & 32 & 4 & 23.46 & 21.93 \\
32 & 1 & 32 & 32 & 19 & 26.66 & 24.80 \\
32 & 1 & 32 & 32 & 32 & 28.96 & 26.46 \\
32 & 1 & 32 & 32 & 188 & 36.70 & 35.29 \\
32 & 1 & 32 & 32 & 313 & 39.06 & 37.67 \\
32 & 1 & 32 & 32 & 1875 & 76.68 & 73.43 \\
32 & 1 & 32 & 32 & 3125 & 109.98 & 106.81 \\
32 & 1 & 32 & 32 & 18750 & 1815.05 & 1688.72 \\
32 & 1 & 32 & 32 & 31250 & 2081.81 & 1979.35 \\
\midrule
      \multicolumn{7}{c}{\mpialltoall $N=1$} \\
32 & 32 & 1 & 32 & 1 & 19.87 & 16.93 \\
32 & 32 & 1 & 32 & 2 & 19.66 & 17.40 \\
32 & 32 & 1 & 32 & 4 & 20.11 & 17.88 \\
32 & 32 & 1 & 32 & 19 & 25.10 & 22.41 \\
32 & 32 & 1 & 32 & 32 & 26.27 & 24.08 \\
32 & 32 & 1 & 32 & 188 & 65.92 & 62.47 \\
32 & 32 & 1 & 32 & 313 & 77.03 & 73.91 \\
32 & 32 & 1 & 32 & 1875 & 145.40 & 141.38 \\
32 & 32 & 1 & 32 & 3125 & 206.43 & 201.46 \\
32 & 32 & 1 & 32 & 18750 & 1775.72 & 1722.57 \\
32 & 32 & 1 & 32 & 31250 & 2848.68 & 2799.03 \\
      \bottomrule
    \end{tabular}
    \end{center}
    \end{small}
\end{table}

For all three MPI libraries, the difference between alltoall on the
node and across nodes is considerable, but also varies greatly between
the libraries, most dramatically for the $k$-ported alltoall
implementations for \openmpiversion where the difference is about a
factor of 10 for larger problems. Also, the simple $k$-ported
implementation with $n=32$ non-blocking MPI send and receive
operations is in most cases (considerably) better for alltoall
communication on a compute note than the MPI library native
\mpialltoall collective.

\clearpage

\subsection{Broadcast}
\label{sec:bcastexperiments}

We compare the $k$-lane and the $k$-ported broadcast implementations
against the full-lane implementation, and the native \mpibcast for the
three different MPI libraries, and try with $k=1,2,3,4,5,6$ virtual
lanes.  The count $c$ is the number of data elements per process. Data
elements are here \mpiint. For the running times (in $\mu$-seconds,
measured with \mpibarrier and \mpiwtime), we report both average and
minimum time of the slowest process over \repetitions repetitions with
\warmup initial, not measured warm-up repetitions.

The broadcast results with \openmpiversion are shown in
Table~\ref{tab:bcast.lane.n32}, Table~\ref{tab:bcast.lane.n32.B},
Table~\ref{tab:bcast.port.n32}, Table~\ref{tab:bcast.port.n32.B} and
Table~\ref{tab:bcast.full.n32}.
The results for the $k$-lane algorithms are disappointing. The
completion overall (for all/most counts) increase with the number of
lanes used for off-node communication. The $k$-ported algorithm is for
all $k$ better than the $k$-lane algorithm, for large counts by a
factor of more than 2. The best algorithm here is the full-lane
algorithm, which outperforms the native \mpibcast by a factor of
about 5 for the largest counts.

\begin{table}
  \caption{Results for $k$-lane Bcast for $k=1,2,3$ on the
    ``Hydra'' system.  The MPI library used is \openmpiversion.}
    \label{tab:bcast.lane.n32}
    \begin{small}
    \begin{center}
    \begin{tabular}{rrrrrrr}
      \toprule
$k$ & $n$ & $N$ & $p$ & $c$ & avg ($\mu s$) & min ($\mu s$) \\
      \midrule
      \multicolumn{7}{c}{Bcast, $k=1$ lane} \\
1 & 32 & 36 & 1152 & 1 & 24.09 & 15.15 \\
1 & 32 & 36 & 1152 & 6 & 24.22 & 17.95 \\
1 & 32 & 36 & 1152 & 10 & 52.51 & 44.42 \\
1 & 32 & 36 & 1152 & 60 & 57.10 & 48.04 \\
1 & 32 & 36 & 1152 & 100 & 33.64 & 25.05 \\
1 & 32 & 36 & 1152 & 600 & 59.89 & 39.32 \\
1 & 32 & 36 & 1152 & 1000 & 46.94 & 43.15 \\
1 & 32 & 36 & 1152 & 6000 & 191.05 & 179.78 \\
1 & 32 & 36 & 1152 & 10000 & 462.90 & 448.33 \\
1 & 32 & 36 & 1152 & 60000 & 2327.44 & 2276.60 \\
1 & 32 & 36 & 1152 & 100000 & 1893.03 & 1770.38 \\
1 & 32 & 36 & 1152 & 600000 & 11822.04 & 10896.21 \\
1 & 32 & 36 & 1152 & 1000000 & 19657.63 & 18363.09 \\
      \midrule
      \multicolumn{7}{c}{Bcast, $k=2$ lanes} \\
2 & 32 & 36 & 1152 & 1 & 19.53 & 15.60 \\
2 & 32 & 36 & 1152 & 6 & 19.92 & 14.70 \\
2 & 32 & 36 & 1152 & 10 & 25.49 & 21.98 \\
2 & 32 & 36 & 1152 & 60 & 26.19 & 22.90 \\
2 & 32 & 36 & 1152 & 100 & 30.44 & 21.51 \\
2 & 32 & 36 & 1152 & 600 & 53.19 & 38.94 \\
2 & 32 & 36 & 1152 & 1000 & 47.82 & 42.64 \\
2 & 32 & 36 & 1152 & 6000 & 206.00 & 187.04 \\
2 & 32 & 36 & 1152 & 10000 & 339.92 & 327.50 \\
2 & 32 & 36 & 1152 & 60000 & 1625.25 & 1571.15 \\
2 & 32 & 36 & 1152 & 100000 & 2592.08 & 2517.84 \\
2 & 32 & 36 & 1152 & 600000 & 16181.73 & 14870.74 \\
2 & 32 & 36 & 1152 & 1000000 & 28057.86 & 26254.84 \\
      \midrule
      \multicolumn{7}{c}{Bcast, $k=3$ lanes} \\
3 & 32 & 36 & 1152 & 1 & 20.86 & 17.17 \\
3 & 32 & 36 & 1152 & 6 & 20.26 & 16.52 \\
3 & 32 & 36 & 1152 & 10 & 26.76 & 23.16 \\
3 & 32 & 36 & 1152 & 60 & 27.89 & 22.24 \\
3 & 32 & 36 & 1152 & 100 & 27.67 & 24.16 \\
3 & 32 & 36 & 1152 & 600 & 56.21 & 44.21 \\
3 & 32 & 36 & 1152 & 1000 & 58.16 & 47.77 \\
3 & 32 & 36 & 1152 & 6000 & 219.06 & 195.99 \\
3 & 32 & 36 & 1152 & 10000 & 330.18 & 317.35 \\
3 & 32 & 36 & 1152 & 60000 & 1594.33 & 1570.48 \\
3 & 32 & 36 & 1152 & 100000 & 3311.97 & 3209.51 \\
3 & 32 & 36 & 1152 & 600000 & 18589.63 & 17941.13 \\
3 & 32 & 36 & 1152 & 1000000 & 31812.81 & 31053.23 \\
      \bottomrule
    \end{tabular}
    \end{center}
    \end{small}
\end{table}

\begin{table}
  \caption{Results for $k$-lane Bcast for $k=4,5,6$ on the
    ``Hydra'' system.  The MPI library used is \openmpiversion.}
    \label{tab:bcast.lane.n32.B}
    \begin{small}
    \begin{center}
    \begin{tabular}{rrrrrrr}
      \toprule
$k$ & $n$ & $N$ & $p$ & $c$ & avg ($\mu s$) & min ($\mu s$) \\
      \midrule
      \multicolumn{7}{c}{Bcast, $k=4$ lanes} \\
4 & 32 & 36 & 1152 & 1 & 17.53 & 14.08 \\
4 & 32 & 36 & 1152 & 6 & 18.58 & 14.58 \\
4 & 32 & 36 & 1152 & 10 & 24.26 & 19.59 \\
4 & 32 & 36 & 1152 & 60 & 24.43 & 20.31 \\
4 & 32 & 36 & 1152 & 100 & 25.97 & 20.84 \\
4 & 32 & 36 & 1152 & 600 & 48.70 & 39.73 \\
4 & 32 & 36 & 1152 & 1000 & 48.25 & 43.70 \\
4 & 32 & 36 & 1152 & 6000 & 195.01 & 185.37 \\
4 & 32 & 36 & 1152 & 10000 & 343.10 & 325.98 \\
4 & 32 & 36 & 1152 & 60000 & 1647.97 & 1621.04 \\
4 & 32 & 36 & 1152 & 100000 & 2673.30 & 2573.66 \\
4 & 32 & 36 & 1152 & 600000 & 15530.99 & 14968.19 \\
4 & 32 & 36 & 1152 & 1000000 & 26839.89 & 25762.55 \\
      \midrule
      \multicolumn{7}{c}{Bcast, $k=5$ lanes} \\
5 & 32 & 36 & 1152 & 1 & 17.47 & 12.89 \\
5 & 32 & 36 & 1152 & 6 & 16.46 & 12.44 \\
5 & 32 & 36 & 1152 & 10 & 22.83 & 17.68 \\
5 & 32 & 36 & 1152 & 60 & 22.97 & 17.79 \\
5 & 32 & 36 & 1152 & 100 & 22.70 & 18.22 \\
5 & 32 & 36 & 1152 & 600 & 46.84 & 37.81 \\
5 & 32 & 36 & 1152 & 1000 & 45.22 & 39.47 \\
5 & 32 & 36 & 1152 & 6000 & 189.61 & 169.03 \\
5 & 32 & 36 & 1152 & 10000 & 278.50 & 248.72 \\
5 & 32 & 36 & 1152 & 60000 & 1285.56 & 1248.00 \\
5 & 32 & 36 & 1152 & 100000 & 2571.22 & 2492.35 \\
5 & 32 & 36 & 1152 & 600000 & 14475.76 & 13944.98 \\
5 & 32 & 36 & 1152 & 1000000 & 25501.07 & 24180.78 \\
      \midrule
      \multicolumn{7}{c}{Bcast, $k=6$ lanes} \\
6 & 32 & 36 & 1152 & 1 & 17.13 & 12.49 \\
6 & 32 & 36 & 1152 & 6 & 16.94 & 13.15 \\
6 & 32 & 36 & 1152 & 10 & 29.59 & 16.48 \\
6 & 32 & 36 & 1152 & 60 & 23.33 & 17.46 \\
6 & 32 & 36 & 1152 & 100 & 25.83 & 17.49 \\
6 & 32 & 36 & 1152 & 600 & 51.22 & 38.15 \\
6 & 32 & 36 & 1152 & 1000 & 53.34 & 41.13 \\
6 & 32 & 36 & 1152 & 6000 & 187.44 & 169.02 \\
6 & 32 & 36 & 1152 & 10000 & 272.23 & 254.72 \\
6 & 32 & 36 & 1152 & 60000 & 1300.32 & 1273.73 \\
6 & 32 & 36 & 1152 & 100000 & 2629.27 & 2509.86 \\
6 & 32 & 36 & 1152 & 600000 & 14708.12 & 14195.74 \\
6 & 32 & 36 & 1152 & 1000000 & 26799.26 & 24599.22 \\
      \bottomrule
    \end{tabular}
    \end{center}
    \end{small}
\end{table}

\begin{table}
  \caption{Results for $k$-ported Bcast for $k=1,2,3$ on the
    ``Hydra'' system.  The MPI library used is \openmpiversion.}
    \label{tab:bcast.port.n32}
\begin{small}
    \begin{center}
    \begin{tabular}{crrrrrrr}
      \toprule
$k$ & $n$ & $N$ & $p$ & $c$ & avg ($\mu s$) & min ($\mu s$) \\
      \midrule
      \multicolumn{7}{c}{Bcast, $1$-ported} \\
1 & 32 & 36 & 1152 & 1 & 15.53 & 11.15 \\
1 & 32 & 36 & 1152 & 6 & 23.44 & 20.80 \\
1 & 32 & 36 & 1152 & 10 & 37.88 & 26.07 \\
1 & 32 & 36 & 1152 & 60 & 30.13 & 25.94 \\
1 & 32 & 36 & 1152 & 100 & 26.39 & 22.04 \\
1 & 32 & 36 & 1152 & 600 & 31.52 & 28.41 \\
1 & 32 & 36 & 1152 & 1000 & 36.07 & 32.08 \\
1 & 32 & 36 & 1152 & 6000 & 139.77 & 133.46 \\
1 & 32 & 36 & 1152 & 10000 & 186.71 & 181.74 \\
1 & 32 & 36 & 1152 & 60000 & 605.62 & 576.52 \\
1 & 32 & 36 & 1152 & 100000 & 943.20 & 906.11 \\
1 & 32 & 36 & 1152 & 600000 & 6283.60 & 5067.77 \\
1 & 32 & 36 & 1152 & 1000000 & 9206.83 & 9014.50 \\
      \midrule
      \multicolumn{7}{c}{Bcast, $2$-ported} \\
2 & 32 & 36 & 1152 & 1 & 18.18 & 9.54 \\
2 & 32 & 36 & 1152 & 6 & 17.46 & 15.07 \\
2 & 32 & 36 & 1152 & 10 & 28.20 & 23.49 \\
2 & 32 & 36 & 1152 & 60 & 28.16 & 20.77 \\
2 & 32 & 36 & 1152 & 100 & 22.35 & 17.22 \\
2 & 32 & 36 & 1152 & 600 & 25.43 & 21.54 \\
2 & 32 & 36 & 1152 & 1000 & 29.14 & 25.44 \\
2 & 32 & 36 & 1152 & 6000 & 109.40 & 102.87 \\
2 & 32 & 36 & 1152 & 10000 & 153.45 & 140.08 \\
2 & 32 & 36 & 1152 & 60000 & 553.37 & 538.79 \\
2 & 32 & 36 & 1152 & 100000 & 893.56 & 855.74 \\
2 & 32 & 36 & 1152 & 600000 & 5444.08 & 5048.19 \\
2 & 32 & 36 & 1152 & 1000000 & 8600.59 & 8256.40 \\
      \midrule
      \multicolumn{7}{c}{Bcast, $3$-ported} \\
3 & 32 & 36 & 1152 & 1 & 14.24 & 9.40 \\
3 & 32 & 36 & 1152 & 6 & 16.47 & 13.89 \\
3 & 32 & 36 & 1152 & 10 & 24.96 & 17.95 \\
3 & 32 & 36 & 1152 & 60 & 30.86 & 18.11 \\
3 & 32 & 36 & 1152 & 100 & 24.35 & 16.04 \\
3 & 32 & 36 & 1152 & 600 & 24.62 & 21.09 \\
3 & 32 & 36 & 1152 & 1000 & 30.01 & 24.35 \\
3 & 32 & 36 & 1152 & 6000 & 109.97 & 96.27 \\
3 & 32 & 36 & 1152 & 10000 & 144.80 & 133.74 \\
3 & 32 & 36 & 1152 & 60000 & 587.78 & 556.06 \\
3 & 32 & 36 & 1152 & 100000 & 950.01 & 893.35 \\
3 & 32 & 36 & 1152 & 600000 & 5745.33 & 5597.42 \\
3 & 32 & 36 & 1152 & 1000000 & 9691.55 & 9298.50 \\
    \bottomrule
    \end{tabular}
    \end{center}
    \end{small}
\end{table}

\begin{table}
  \caption{Results for $k$-ported Bcast for $k=4,5,6$ on the
    ``Hydra'' system.  The MPI library used is \openmpiversion.}
    \label{tab:bcast.port.n32.B}
\begin{small}
    \begin{center}
    \begin{tabular}{rrrrrrr}
      \toprule
$k$ & $n$ & $N$ & $p$ & $c$ & avg ($\mu s$) & min ($\mu s$) \\
      \midrule
      \multicolumn{7}{c}{Bcast, $4$-ported} \\
4 & 32 & 36 & 1152 & 1 & 13.82 & 9.00 \\
4 & 32 & 36 & 1152 & 6 & 15.83 & 12.28 \\
4 & 32 & 36 & 1152 & 10 & 23.55 & 17.66 \\
4 & 32 & 36 & 1152 & 60 & 33.11 & 19.98 \\
4 & 32 & 36 & 1152 & 100 & 19.79 & 14.60 \\
4 & 32 & 36 & 1152 & 600 & 22.70 & 18.57 \\
4 & 32 & 36 & 1152 & 1000 & 27.67 & 23.57 \\
4 & 32 & 36 & 1152 & 6000 & 103.03 & 94.45 \\
4 & 32 & 36 & 1152 & 10000 & 135.18 & 125.68 \\
4 & 32 & 36 & 1152 & 60000 & 619.11 & 588.26 \\
4 & 32 & 36 & 1152 & 100000 & 971.62 & 952.70 \\
4 & 32 & 36 & 1152 & 600000 & 5739.24 & 5561.88 \\
4 & 32 & 36 & 1152 & 1000000 & 9771.26 & 9044.56 \\
      \midrule
      \multicolumn{7}{c}{Bcast, $5$-ported} \\
5 & 32 & 36 & 1152 & 1 & 14.83 & 9.69 \\
5 & 32 & 36 & 1152 & 6 & 14.55 & 11.16 \\
5 & 32 & 36 & 1152 & 10 & 30.30 & 23.53 \\
5 & 32 & 36 & 1152 & 60 & 35.14 & 22.92 \\
5 & 32 & 36 & 1152 & 100 & 45.81 & 15.04 \\
5 & 32 & 36 & 1152 & 600 & 26.38 & 19.24 \\
5 & 32 & 36 & 1152 & 1000 & 30.86 & 23.77 \\
5 & 32 & 36 & 1152 & 6000 & 121.11 & 93.29 \\
5 & 32 & 36 & 1152 & 10000 & 144.91 & 128.41 \\
5 & 32 & 36 & 1152 & 60000 & 725.90 & 582.30 \\
5 & 32 & 36 & 1152 & 100000 & 982.30 & 953.40 \\
5 & 32 & 36 & 1152 & 600000 & 5886.35 & 5729.35 \\
5 & 32 & 36 & 1152 & 1000000 & 9965.33 & 9698.53 \\
      \midrule
      \multicolumn{7}{c}{Bcast, $6$-ported} \\
6 & 32 & 36 & 1152 & 1 & 14.85 & 9.67 \\
6 & 32 & 36 & 1152 & 6 & 16.26 & 12.56 \\
6 & 32 & 36 & 1152 & 10 & 24.47 & 20.00 \\
6 & 32 & 36 & 1152 & 60 & 24.74 & 19.42 \\
6 & 32 & 36 & 1152 & 100 & 24.88 & 15.58 \\
6 & 32 & 36 & 1152 & 600 & 23.03 & 19.78 \\
6 & 32 & 36 & 1152 & 1000 & 28.17 & 24.38 \\
6 & 32 & 36 & 1152 & 6000 & 102.05 & 95.98 \\
6 & 32 & 36 & 1152 & 10000 & 136.73 & 128.96 \\
6 & 32 & 36 & 1152 & 60000 & 649.85 & 610.34 \\
6 & 32 & 36 & 1152 & 100000 & 1056.39 & 1014.92 \\
6 & 32 & 36 & 1152 & 600000 & 6253.67 & 5936.32 \\
6 & 32 & 36 & 1152 & 1000000 & 10819.07 & 10224.51 \\
    \bottomrule
    \end{tabular}
    \end{center}
    \end{small}
\end{table}

\begin{table}
  \caption{Results for full-lane Bcast and the native \mpibcast on the
    ``Hydra'' system.  The MPI library used is \openmpiversion.}
    \label{tab:bcast.full.n32}
    \begin{small}
    \begin{center}
    \begin{tabular}{crrrrrrr}
      \toprule
$k$ & $n$ & $N$ & $p$ & $c$ & avg ($\mu s$) & min ($\mu s$) \\
      \midrule
      \multicolumn{7}{c}{Full-lane Bcast} \\
6 & 32 & 36 & 1152 & 1 & 19.84 & 13.99 \\
6 & 32 & 36 & 1152 & 6 & 20.84 & 15.63 \\
6 & 32 & 36 & 1152 & 10 & 38.42 & 21.39 \\
6 & 32 & 36 & 1152 & 60 & 50.70 & 30.00 \\
6 & 32 & 36 & 1152 & 100 & 40.31 & 28.71 \\
6 & 32 & 36 & 1152 & 600 & 71.78 & 43.44 \\
6 & 32 & 36 & 1152 & 1000 & 48.90 & 43.08 \\
6 & 32 & 36 & 1152 & 6000 & 70.74 & 55.05 \\
6 & 32 & 36 & 1152 & 10000 & 82.44 & 73.80 \\
6 & 32 & 36 & 1152 & 60000 & 205.95 & 190.59 \\
6 & 32 & 36 & 1152 & 100000 & 302.71 & 278.23 \\
6 & 32 & 36 & 1152 & 600000 & 1936.82 & 1840.89 \\
6 & 32 & 36 & 1152 & 1000000 & 3309.16 & 3244.24 \\
\midrule
\multicolumn{7}{c}{\mpibcast} \\
6 & 32 & 36 & 1152 & 1 & 13.44 & 9.11 \\
6 & 32 & 36 & 1152 & 6 & 15.26 & 11.50 \\
6 & 32 & 36 & 1152 & 10 & 15.84 & 12.40 \\
6 & 32 & 36 & 1152 & 60 & 16.48 & 12.25 \\
6 & 32 & 36 & 1152 & 100 & 16.91 & 14.00 \\
6 & 32 & 36 & 1152 & 600 & 31.36 & 27.02 \\
6 & 32 & 36 & 1152 & 1000 & 34.50 & 30.27 \\
6 & 32 & 36 & 1152 & 6000 & 93.90 & 81.58 \\
6 & 32 & 36 & 1152 & 10000 & 128.91 & 117.64 \\
6 & 32 & 36 & 1152 & 60000 & 642.72 & 585.93 \\
6 & 32 & 36 & 1152 & 100000 & 8753.50 & 7651.04 \\
6 & 32 & 36 & 1152 & 600000 & 15598.81 & 13772.96 \\
6 & 32 & 36 & 1152 & 1000000 & 18067.27 & 17081.58 \\
    \bottomrule
    \end{tabular}
    \end{center}
    \end{small}
\end{table}

The broadcast results with \intelmpiversion are shown in
Table~\ref{tab:bcast.lane.n32.intel},
Table~\ref{tab:bcast.lane.n32.intel.B},
Table~\ref{tab:bcast.port.n32.intel},
Table~\ref{tab:bcast.port.n32.intel.B} and
Table~\ref{tab:bcast.full.n32.intel}. The results are qualitatively
similar to those for the \openmpiversion library, with again the
full-lane algorithm being considerably better than \mpibcast. For
\intelmpiversion, \mpibcast is terrible for small $c$, and needs to be
repaired or tuned better (algorithm selection).

\begin{table}
  \caption{Results for $k$-lane Bcast for $k=1,2,3$ on the
    ``Hydra'' system.  The MPI library used is \intelmpiversion.}
    \label{tab:bcast.lane.n32.intel}
    \begin{small}
    \begin{center}
    \begin{tabular}{rrrrrrr}
      \toprule
$k$ & $n$ & $N$ & $p$ & $c$ & avg ($\mu s$) & min ($\mu s$) \\
      \midrule
      \multicolumn{7}{c}{Bcast, $1$ lane} \\
1 & 32 & 36 & 1152 & 1 & 37.45 & 30.99 \\
1 & 32 & 36 & 1152 & 6 & 35.42 & 31.95 \\
1 & 32 & 36 & 1152 & 10 & 35.47 & 31.95 \\
1 & 32 & 36 & 1152 & 60 & 48.87 & 45.06 \\
1 & 32 & 36 & 1152 & 100 & 52.03 & 47.92 \\
1 & 32 & 36 & 1152 & 600 & 81.86 & 78.92 \\
1 & 32 & 36 & 1152 & 1000 & 105.33 & 100.85 \\
1 & 32 & 36 & 1152 & 6000 & 316.04 & 281.10 \\
1 & 32 & 36 & 1152 & 10000 & 344.60 & 315.90 \\
1 & 32 & 36 & 1152 & 60000 & 869.94 & 813.01 \\
1 & 32 & 36 & 1152 & 100000 & 1265.99 & 1240.97 \\
1 & 32 & 36 & 1152 & 600000 & 8419.50 & 8136.03 \\
1 & 32 & 36 & 1152 & 1000000 & 19502.91 & 18983.13 \\
      \midrule
      \multicolumn{7}{c}{Bcast, $2$ lanes} \\
2 & 32 & 36 & 1152 & 1 & 58.95 & 56.03 \\
2 & 32 & 36 & 1152 & 6 & 59.98 & 56.98 \\
2 & 32 & 36 & 1152 & 10 & 59.55 & 56.03 \\
2 & 32 & 36 & 1152 & 60 & 75.88 & 72.00 \\
2 & 32 & 36 & 1152 & 100 & 83.70 & 79.15 \\
2 & 32 & 36 & 1152 & 600 & 142.00 & 133.04 \\
2 & 32 & 36 & 1152 & 1000 & 179.55 & 175.00 \\
2 & 32 & 36 & 1152 & 6000 & 433.68 & 422.95 \\
2 & 32 & 36 & 1152 & 10000 & 510.23 & 489.00 \\
2 & 32 & 36 & 1152 & 60000 & 1105.82 & 1086.00 \\
2 & 32 & 36 & 1152 & 100000 & 1762.17 & 1733.06 \\
2 & 32 & 36 & 1152 & 600000 & 11456.29 & 10638.95 \\
2 & 32 & 36 & 1152 & 1000000 & 19808.21 & 19363.88 \\
      \midrule
      \multicolumn{7}{c}{Bcast, $3$ lanes} \\
3 & 32 & 36 & 1152 & 1 & 117.68 & 69.14 \\
3 & 32 & 36 & 1152 & 6 & 82.05 & 76.06 \\
3 & 32 & 36 & 1152 & 10 & 80.92 & 77.01 \\
3 & 32 & 36 & 1152 & 60 & 102.78 & 97.04 \\
3 & 32 & 36 & 1152 & 100 & 113.36 & 106.81 \\
3 & 32 & 36 & 1152 & 600 & 193.63 & 185.01 \\
3 & 32 & 36 & 1152 & 1000 & 251.56 & 243.90 \\
3 & 32 & 36 & 1152 & 6000 & 598.35 & 586.03 \\
3 & 32 & 36 & 1152 & 10000 & 684.79 & 660.18 \\
3 & 32 & 36 & 1152 & 60000 & 1369.17 & 1346.11 \\
3 & 32 & 36 & 1152 & 100000 & 2096.83 & 2069.95 \\
3 & 32 & 36 & 1152 & 600000 & 11982.61 & 11868.00 \\
3 & 32 & 36 & 1152 & 1000000 & 20510.85 & 19860.03 \\
    \bottomrule
    \end{tabular}
    \end{center}
    \end{small}
\end{table}

\begin{table}
  \caption{Results for $k$-lane Bcast for $k=4,5,6$ on the
    ``Hydra'' system.  The MPI library used is \intelmpiversion.}
    \label{tab:bcast.lane.n32.intel.B}
    \begin{small}
    \begin{center}
    \begin{tabular}{rrrrrrr}
      \toprule
$k$ & $n$ & $N$ & $p$ & $c$ & avg ($\mu s$) & min ($\mu s$) \\
      \midrule
      \multicolumn{7}{c}{Bcast, $4$ lanes} \\
4 & 32 & 36 & 1152 & 1 & 59.04 & 55.07 \\
4 & 32 & 36 & 1152 & 6 & 59.52 & 56.03 \\
4 & 32 & 36 & 1152 & 10 & 59.00 & 55.07 \\
4 & 32 & 36 & 1152 & 60 & 75.01 & 72.00 \\
4 & 32 & 36 & 1152 & 100 & 82.08 & 77.96 \\
4 & 32 & 36 & 1152 & 600 & 134.87 & 130.89 \\
4 & 32 & 36 & 1152 & 1000 & 176.19 & 172.14 \\
4 & 32 & 36 & 1152 & 6000 & 425.08 & 416.99 \\
4 & 32 & 36 & 1152 & 10000 & 489.01 & 474.93 \\
4 & 32 & 36 & 1152 & 60000 & 1064.72 & 1042.13 \\
4 & 32 & 36 & 1152 & 100000 & 1681.76 & 1651.05 \\
4 & 32 & 36 & 1152 & 600000 & 10364.62 & 10273.93 \\
4 & 32 & 36 & 1152 & 1000000 & 18577.58 & 18381.12 \\
      \midrule
      \multicolumn{7}{c}{Bcast, $5$ lanes} \\
5 & 32 & 36 & 1152 & 1 & 59.61 & 56.03 \\
5 & 32 & 36 & 1152 & 6 & 59.98 & 56.03 \\
5 & 32 & 36 & 1152 & 10 & 58.73 & 55.07 \\
5 & 32 & 36 & 1152 & 60 & 76.68 & 70.81 \\
5 & 32 & 36 & 1152 & 100 & 187.29 & 78.20 \\
5 & 32 & 36 & 1152 & 600 & 143.55 & 128.98 \\
5 & 32 & 36 & 1152 & 1000 & 179.49 & 174.05 \\
5 & 32 & 36 & 1152 & 6000 & 427.26 & 416.04 \\
5 & 32 & 36 & 1152 & 10000 & 481.25 & 469.21 \\
5 & 32 & 36 & 1152 & 60000 & 1007.28 & 986.81 \\
5 & 32 & 36 & 1152 & 100000 & 1571.63 & 1544.95 \\
5 & 32 & 36 & 1152 & 600000 & 9268.71 & 9174.11 \\
5 & 32 & 36 & 1152 & 1000000 & 16099.62 & 15856.03 \\
      \midrule
      \multicolumn{7}{c}{Bcast, $6$ lanes} \\
6 & 32 & 36 & 1152 & 1 & 59.68 & 56.98 \\
6 & 32 & 36 & 1152 & 6 & 59.75 & 56.03 \\
6 & 32 & 36 & 1152 & 10 & 59.33 & 56.03 \\
6 & 32 & 36 & 1152 & 60 & 75.46 & 71.05 \\
6 & 32 & 36 & 1152 & 100 & 103.45 & 77.96 \\
6 & 32 & 36 & 1152 & 600 & 147.82 & 133.99 \\
6 & 32 & 36 & 1152 & 1000 & 191.08 & 175.00 \\
6 & 32 & 36 & 1152 & 6000 & 427.77 & 417.95 \\
6 & 32 & 36 & 1152 & 10000 & 481.02 & 468.97 \\
6 & 32 & 36 & 1152 & 60000 & 1026.83 & 987.05 \\
6 & 32 & 36 & 1152 & 100000 & 1586.57 & 1564.98 \\
6 & 32 & 36 & 1152 & 600000 & 9515.10 & 9393.93 \\
6 & 32 & 36 & 1152 & 1000000 & 16714.83 & 16155.96 \\
    \bottomrule
    \end{tabular}
    \end{center}
    \end{small}
\end{table}

\begin{table}
  \caption{Results for $k$-ported Bcast for $k=1,2,3$ on the
    ``Hydra'' system.  The MPI library used is \intelmpiversion.}
    \label{tab:bcast.port.n32.intel}
    \begin{small}
    \begin{center}
    \begin{tabular}{rrrrrrr}
      \toprule
$k$ & $n$ & $N$ & $p$ & $c$ & avg ($\mu s$) & min ($\mu s$) \\
      \midrule
      \multicolumn{7}{c}{Bcast, $1$-ported} \\
1 & 32 & 36 & 1152 & 1 & 13.88 & 9.06 \\
1 & 32 & 36 & 1152 & 6 & 13.81 & 10.01 \\
1 & 32 & 36 & 1152 & 10 & 13.83 & 10.01 \\
1 & 32 & 36 & 1152 & 60 & 24.77 & 22.89 \\
1 & 32 & 36 & 1152 & 100 & 25.76 & 22.89 \\
1 & 32 & 36 & 1152 & 600 & 35.22 & 29.09 \\
1 & 32 & 36 & 1152 & 1000 & 35.98 & 34.09 \\
1 & 32 & 36 & 1152 & 6000 & 103.94 & 100.85 \\
1 & 32 & 36 & 1152 & 10000 & 131.42 & 126.12 \\
1 & 32 & 36 & 1152 & 60000 & 589.08 & 579.12 \\
1 & 32 & 36 & 1152 & 100000 & 898.37 & 880.00 \\
1 & 32 & 36 & 1152 & 600000 & 5718.99 & 5054.95 \\
1 & 32 & 36 & 1152 & 1000000 & 9342.87 & 9306.91 \\
      \midrule
      \multicolumn{7}{c}{Bcast, $2$-ported} \\
2 & 32 & 36 & 1152 & 1 & 13.16 & 8.82 \\
2 & 32 & 36 & 1152 & 6 & 12.69 & 9.06 \\
2 & 32 & 36 & 1152 & 10 & 12.00 & 9.06 \\
2 & 32 & 36 & 1152 & 60 & 17.74 & 15.97 \\
2 & 32 & 36 & 1152 & 100 & 17.96 & 15.97 \\
2 & 32 & 36 & 1152 & 600 & 23.19 & 20.98 \\
2 & 32 & 36 & 1152 & 1000 & 27.05 & 25.03 \\
2 & 32 & 36 & 1152 & 6000 & 88.28 & 86.07 \\
2 & 32 & 36 & 1152 & 10000 & 118.23 & 114.92 \\
2 & 32 & 36 & 1152 & 60000 & 558.77 & 539.06 \\
2 & 32 & 36 & 1152 & 100000 & 866.13 & 845.91 \\
2 & 32 & 36 & 1152 & 600000 & 4985.75 & 4918.10 \\
2 & 32 & 36 & 1152 & 1000000 & 8174.66 & 8090.02 \\
      \midrule
      \multicolumn{7}{c}{Bcast, $3$-ported} \\
3 & 32 & 36 & 1152 & 1 & 11.76 & 8.11 \\
3 & 32 & 36 & 1152 & 6 & 12.03 & 9.06 \\
3 & 32 & 36 & 1152 & 10 & 12.25 & 9.06 \\
3 & 32 & 36 & 1152 & 60 & 15.75 & 14.07 \\
3 & 32 & 36 & 1152 & 100 & 15.73 & 14.07 \\
3 & 32 & 36 & 1152 & 600 & 21.81 & 18.12 \\
3 & 32 & 36 & 1152 & 1000 & 25.28 & 22.89 \\
3 & 32 & 36 & 1152 & 6000 & 87.95 & 85.12 \\
3 & 32 & 36 & 1152 & 10000 & 114.62 & 111.10 \\
3 & 32 & 36 & 1152 & 60000 & 568.79 & 553.13 \\
3 & 32 & 36 & 1152 & 100000 & 908.80 & 894.07 \\
3 & 32 & 36 & 1152 & 600000 & 5494.14 & 5321.03 \\
3 & 32 & 36 & 1152 & 1000000 & 9380.43 & 9196.04 \\
    \bottomrule
    \end{tabular}
    \end{center}
    \end{small}
\end{table}

\begin{table}
  \caption{Results for $k$-ported Bcast for $k=4,5,6$ on the
    ``Hydra'' system.  The MPI library used is \intelmpiversion.}
    \label{tab:bcast.port.n32.intel.B}
    \begin{small}
    \begin{center}
    \begin{tabular}{rrrrrrr}
      \toprule
$k$ & $n$ & $N$ & $p$ & $c$ & avg ($\mu s$) & min ($\mu s$) \\
      \midrule
      \multicolumn{7}{c}{Bcast, $4$-ported} \\
4 & 32 & 36 & 1152 & 1 & 11.85 & 7.87 \\
4 & 32 & 36 & 1152 & 6 & 12.41 & 10.01 \\
4 & 32 & 36 & 1152 & 10 & 11.84 & 9.78 \\
4 & 32 & 36 & 1152 & 60 & 16.47 & 13.11 \\
4 & 32 & 36 & 1152 & 100 & 16.26 & 14.07 \\
4 & 32 & 36 & 1152 & 600 & 20.88 & 18.12 \\
4 & 32 & 36 & 1152 & 1000 & 24.68 & 21.93 \\
4 & 32 & 36 & 1152 & 6000 & 87.93 & 83.92 \\
4 & 32 & 36 & 1152 & 10000 & 124.62 & 106.10 \\
4 & 32 & 36 & 1152 & 60000 & 568.49 & 539.06 \\
4 & 32 & 36 & 1152 & 100000 & 919.77 & 872.14 \\
4 & 32 & 36 & 1152 & 600000 & 5421.76 & 5332.95 \\
4 & 32 & 36 & 1152 & 1000000 & 9146.91 & 9001.97 \\
      \midrule
      \multicolumn{7}{c}{Bcast, $5$-ported} \\
5 & 32 & 36 & 1152 & 1 & 11.20 & 7.87 \\
5 & 32 & 36 & 1152 & 6 & 21.66 & 8.11 \\
5 & 32 & 36 & 1152 & 10 & 45.39 & 8.82 \\
5 & 32 & 36 & 1152 & 60 & 24.62 & 10.97 \\
5 & 32 & 36 & 1152 & 100 & 21.87 & 11.92 \\
5 & 32 & 36 & 1152 & 600 & 22.92 & 16.93 \\
5 & 32 & 36 & 1152 & 1000 & 26.10 & 21.93 \\
5 & 32 & 36 & 1152 & 6000 & 86.49 & 82.97 \\
5 & 32 & 36 & 1152 & 10000 & 117.03 & 111.10 \\
5 & 32 & 36 & 1152 & 60000 & 614.95 & 596.05 \\
5 & 32 & 36 & 1152 & 100000 & 987.40 & 957.01 \\
5 & 32 & 36 & 1152 & 600000 & 5746.27 & 5664.11 \\
5 & 32 & 36 & 1152 & 1000000 & 9835.49 & 9638.07 \\
      \midrule
      \multicolumn{7}{c}{Bcast, $6$-ported} \\
6 & 32 & 36 & 1152 & 1 & 12.06 & 8.82 \\
6 & 32 & 36 & 1152 & 6 & 11.60 & 10.01 \\
6 & 32 & 36 & 1152 & 10 & 11.96 & 10.01 \\
6 & 32 & 36 & 1152 & 60 & 14.48 & 12.16 \\
6 & 32 & 36 & 1152 & 100 & 15.04 & 13.11 \\
6 & 32 & 36 & 1152 & 600 & 20.23 & 18.12 \\
6 & 32 & 36 & 1152 & 1000 & 25.11 & 22.89 \\
6 & 32 & 36 & 1152 & 6000 & 99.28 & 94.18 \\
6 & 32 & 36 & 1152 & 10000 & 128.16 & 125.17 \\
6 & 32 & 36 & 1152 & 60000 & 608.25 & 586.99 \\
6 & 32 & 36 & 1152 & 100000 & 993.33 & 959.87 \\
6 & 32 & 36 & 1152 & 600000 & 5904.18 & 5725.15 \\
6 & 32 & 36 & 1152 & 1000000 & 10360.80 & 10004.04 \\
    \bottomrule
    \end{tabular}
    \end{center}
    \end{small}
\end{table}

\begin{table}
  \caption{Results for full-lane Bcast and the native \mpibcast on the
    ``Hydra'' system.  The MPI library used is \intelmpiversion.}
    \label{tab:bcast.full.n32.intel}
    \begin{small}
    \begin{center}
    \begin{tabular}{rrrrrrr}
      \toprule
$k$ & $n$ & $N$ & $p$ & $c$ & avg ($\mu s$) & min ($\mu s$) \\
      \midrule
      \multicolumn{7}{c}{Full-lane Bcast} \\
6 & 32 & 36 & 1152 & 1 & 57.94 & 51.98 \\
6 & 32 & 36 & 1152 & 6 & 64.52 & 61.04 \\
6 & 32 & 36 & 1152 & 10 & 64.17 & 61.04 \\
6 & 32 & 36 & 1152 & 60 & 81.45 & 78.92 \\
6 & 32 & 36 & 1152 & 100 & 77.16 & 72.96 \\
6 & 32 & 36 & 1152 & 600 & 83.63 & 79.87 \\
6 & 32 & 36 & 1152 & 1000 & 86.52 & 82.02 \\
6 & 32 & 36 & 1152 & 6000 & 102.75 & 97.99 \\
6 & 32 & 36 & 1152 & 10000 & 115.68 & 107.05 \\
6 & 32 & 36 & 1152 & 60000 & 466.27 & 440.84 \\
6 & 32 & 36 & 1152 & 100000 & 662.22 & 644.92 \\
6 & 32 & 36 & 1152 & 600000 & 2755.01 & 2590.18 \\
6 & 32 & 36 & 1152 & 1000000 & 4268.80 & 4073.86 \\
      \midrule
      \multicolumn{7}{c}{\mpibcast} \\
6 & 32 & 36 & 1152 & 1 & 965.34 & 933.89 \\
6 & 32 & 36 & 1152 & 6 & 980.15 & 964.88 \\
6 & 32 & 36 & 1152 & 10 & 987.97 & 971.08 \\
6 & 32 & 36 & 1152 & 60 & 1258.53 & 1237.15 \\
6 & 32 & 36 & 1152 & 100 & 1392.92 & 1369.95 \\
6 & 32 & 36 & 1152 & 600 & 2290.98 & 2268.08 \\
6 & 32 & 36 & 1152 & 1000 & 3155.77 & 2979.99 \\
6 & 32 & 36 & 1152 & 6000 & 6900.65 & 6588.94 \\
6 & 32 & 36 & 1152 & 10000 & 6820.39 & 6774.90 \\
6 & 32 & 36 & 1152 & 60000 & 7992.54 & 7817.98 \\
6 & 32 & 36 & 1152 & 100000 & 8504.42 & 8333.21 \\
6 & 32 & 36 & 1152 & 600000 & 13562.95 & 13094.19 \\
6 & 32 & 36 & 1152 & 1000000 & 16058.13 & 15733.96 \\
    \bottomrule
    \end{tabular}
    \end{center}
    \end{small}
\end{table}

The broadcast results with \mpichversion are shown in
Table~\ref{tab:bcast.lane.n32.mpich},
Table~\ref{tab:bcast.lane.n32.mpich.B},
Table~\ref{tab:bcast.port.n32.mpich},
Table~\ref{tab:bcast.port.n32.mpich.B} and
Table~\ref{tab:bcast.full.n32.mpich}.
From the $k$-lane algorithm perspective this MPI library gives the
most interesting performance results. The running time decrease with
increasing $k$, with overall best performance for $k=5,6$. In absolute
terms, the $k$-ported algorithm performs much better, though, for
large counts, but performance seems largely independent of the chosen
$k$. Also here, the full-lane implementations performs the best, but
the difference to the native \mpibcast is smaller than for the other
libraries. \mpibcast is by far the best for small $c$.

\begin{table}
  \caption{Results for $k$-lane Bcast for $k=1,2,3$ on the
    ``Hydra'' system.  The MPI library used is \mpichversion.}
    \label{tab:bcast.lane.n32.mpich}
    \begin{small}
    \begin{center}
    \begin{tabular}{rrrrrrr}
      \toprule
$k$ & $n$ & $N$ & $p$ & $c$ & avg ($\mu s$) & min ($\mu s$) \\
\midrule
\multicolumn{7}{c}{Bcast, $1$ lane} \\
1 & 32 & 36 & 1152 & 1 & 19.11 & 14.07 \\
1 & 32 & 36 & 1152 & 6 & 19.64 & 15.97 \\
1 & 32 & 36 & 1152 & 10 & 19.43 & 15.74 \\
1 & 32 & 36 & 1152 & 60 & 106.60 & 20.98 \\
1 & 32 & 36 & 1152 & 100 & 76.61 & 20.50 \\
1 & 32 & 36 & 1152 & 600 & 66.81 & 40.05 \\
1 & 32 & 36 & 1152 & 1000 & 51.18 & 44.82 \\
1 & 32 & 36 & 1152 & 6000 & 452.14 & 272.99 \\
1 & 32 & 36 & 1152 & 10000 & 472.84 & 325.92 \\
1 & 32 & 36 & 1152 & 60000 & 1720.74 & 982.52 \\
1 & 32 & 36 & 1152 & 100000 & 1571.35 & 1476.76 \\
1 & 32 & 36 & 1152 & 600000 & 16047.56 & 15645.98 \\
1 & 32 & 36 & 1152 & 1000000 & 26739.39 & 26161.91 \\
\midrule
\multicolumn{7}{c}{Bcast, $2$ lanes} \\
2 & 32 & 36 & 1152 & 1 & 20.59 & 16.69 \\
2 & 32 & 36 & 1152 & 6 & 20.60 & 15.74 \\
2 & 32 & 36 & 1152 & 10 & 20.32 & 16.45 \\
2 & 32 & 36 & 1152 & 60 & 22.12 & 17.88 \\
2 & 32 & 36 & 1152 & 100 & 23.12 & 19.31 \\
2 & 32 & 36 & 1152 & 600 & 36.75 & 33.86 \\
2 & 32 & 36 & 1152 & 1000 & 48.92 & 42.92 \\
2 & 32 & 36 & 1152 & 6000 & 206.84 & 193.60 \\
2 & 32 & 36 & 1152 & 10000 & 257.26 & 239.37 \\
2 & 32 & 36 & 1152 & 60000 & 824.01 & 766.52 \\
2 & 32 & 36 & 1152 & 100000 & 1255.18 & 1198.29 \\
2 & 32 & 36 & 1152 & 600000 & 12249.17 & 11814.83 \\
2 & 32 & 36 & 1152 & 1000000 & 20230.43 & 19976.14 \\
\midrule
\multicolumn{7}{c}{Bcast, $3$ lanes} \\
3 & 32 & 36 & 1152 & 1 & 18.82 & 15.26 \\
3 & 32 & 36 & 1152 & 6 & 18.42 & 14.78 \\
3 & 32 & 36 & 1152 & 10 & 19.11 & 15.50 \\
3 & 32 & 36 & 1152 & 60 & 20.34 & 17.17 \\
3 & 32 & 36 & 1152 & 100 & 21.53 & 18.12 \\
3 & 32 & 36 & 1152 & 600 & 34.06 & 29.80 \\
3 & 32 & 36 & 1152 & 1000 & 44.71 & 41.01 \\
3 & 32 & 36 & 1152 & 6000 & 201.19 & 184.30 \\
3 & 32 & 36 & 1152 & 10000 & 249.38 & 231.74 \\
3 & 32 & 36 & 1152 & 60000 & 801.17 & 736.95 \\
3 & 32 & 36 & 1152 & 100000 & 1217.80 & 1169.20 \\
3 & 32 & 36 & 1152 & 600000 & 12127.84 & 11790.99 \\
3 & 32 & 36 & 1152 & 1000000 & 19952.22 & 19467.83 \\
    \bottomrule
    \end{tabular}
    \end{center}
    \end{small}
\end{table}

\begin{table}
  \caption{Results for $k$-lane Bcast for $k=4,5,6$ on the
    ``Hydra'' system.  The MPI library used is \mpichversion.}
    \label{tab:bcast.lane.n32.mpich.B}
    \begin{small}
    \begin{center}
    \begin{tabular}{rrrrrrr}
      \toprule
$k$ & $n$ & $N$ & $p$ & $c$ & avg ($\mu s$) & min ($\mu s$) \\
      \midrule
      \multicolumn{7}{c}{Bcast, $4$ lanes} \\
4 & 32 & 36 & 1152 & 1 & 21.49 & 13.59 \\
4 & 32 & 36 & 1152 & 6 & 17.82 & 14.54 \\
4 & 32 & 36 & 1152 & 10 & 17.94 & 14.78 \\
4 & 32 & 36 & 1152 & 60 & 18.94 & 15.74 \\
4 & 32 & 36 & 1152 & 100 & 19.23 & 16.69 \\
4 & 32 & 36 & 1152 & 600 & 34.69 & 27.18 \\
4 & 32 & 36 & 1152 & 1000 & 43.59 & 38.15 \\
4 & 32 & 36 & 1152 & 6000 & 193.46 & 175.00 \\
4 & 32 & 36 & 1152 & 10000 & 232.99 & 220.54 \\
4 & 32 & 36 & 1152 & 60000 & 761.45 & 715.02 \\
4 & 32 & 36 & 1152 & 100000 & 1173.01 & 1119.38 \\
4 & 32 & 36 & 1152 & 600000 & 11718.44 & 11438.85 \\
4 & 32 & 36 & 1152 & 1000000 & 19136.06 & 18922.33 \\
      \midrule
      \multicolumn{7}{c}{Bcast, $5$ lanes} \\
5 & 32 & 36 & 1152 & 1 & 17.59 & 13.59 \\
5 & 32 & 36 & 1152 & 6 & 17.31 & 14.07 \\
5 & 32 & 36 & 1152 & 10 & 16.96 & 12.87 \\
5 & 32 & 36 & 1152 & 60 & 18.36 & 15.26 \\
5 & 32 & 36 & 1152 & 100 & 19.13 & 15.50 \\
5 & 32 & 36 & 1152 & 600 & 33.61 & 27.66 \\
5 & 32 & 36 & 1152 & 1000 & 40.73 & 36.24 \\
5 & 32 & 36 & 1152 & 6000 & 155.93 & 143.29 \\
5 & 32 & 36 & 1152 & 10000 & 197.90 & 184.30 \\
5 & 32 & 36 & 1152 & 60000 & 684.09 & 611.31 \\
5 & 32 & 36 & 1152 & 100000 & 1035.46 & 983.24 \\
5 & 32 & 36 & 1152 & 600000 & 9689.02 & 9536.98 \\
5 & 32 & 36 & 1152 & 1000000 & 16199.05 & 15951.87 \\
      \midrule
      \multicolumn{7}{c}{Bcast, $6$ lanes} \\
6 & 32 & 36 & 1152 & 1 & 20.30 & 14.31 \\
6 & 32 & 36 & 1152 & 6 & 18.04 & 14.31 \\
6 & 32 & 36 & 1152 & 10 & 21.95 & 14.54 \\
6 & 32 & 36 & 1152 & 60 & 17.95 & 15.26 \\
6 & 32 & 36 & 1152 & 100 & 20.91 & 15.97 \\
6 & 32 & 36 & 1152 & 600 & 37.96 & 28.85 \\
6 & 32 & 36 & 1152 & 1000 & 41.50 & 38.15 \\
6 & 32 & 36 & 1152 & 6000 & 163.11 & 144.48 \\
6 & 32 & 36 & 1152 & 10000 & 198.47 & 182.39 \\
6 & 32 & 36 & 1152 & 60000 & 672.31 & 625.61 \\
6 & 32 & 36 & 1152 & 100000 & 1091.53 & 984.91 \\
6 & 32 & 36 & 1152 & 600000 & 9700.79 & 9450.91 \\
6 & 32 & 36 & 1152 & 1000000 & 15891.74 & 15699.86 \\
    \bottomrule
    \end{tabular}
    \end{center}
    \end{small}
\end{table}

\begin{table}
  \caption{Results for $k$-ported Bcast for $k=1,2,3$ on the
    ``Hydra'' system.  The MPI library used is \mpichversion.}
    \label{tab:bcast.port.n32.mpich}
    \begin{small}
    \begin{center}
    \begin{tabular}{rrrrrrr}
      \toprule
$k$ & $n$ & $N$ & $p$ & $c$ & avg ($\mu s$) & min ($\mu s$) \\
      \midrule
      \multicolumn{7}{c}{Bcast, $1$-ported} \\
1 & 32 & 36 & 1152 & 1 & 20.76 & 13.35 \\
1 & 32 & 36 & 1152 & 6 & 17.97 & 13.83 \\
1 & 32 & 36 & 1152 & 10 & 15.85 & 11.68 \\
1 & 32 & 36 & 1152 & 60 & 31.89 & 25.99 \\
1 & 32 & 36 & 1152 & 100 & 34.11 & 26.46 \\
1 & 32 & 36 & 1152 & 600 & 46.02 & 34.09 \\
1 & 32 & 36 & 1152 & 1000 & 61.32 & 36.72 \\
1 & 32 & 36 & 1152 & 6000 & 268.57 & 118.97 \\
1 & 32 & 36 & 1152 & 10000 & 424.39 & 153.30 \\
1 & 32 & 36 & 1152 & 60000 & 585.59 & 454.19 \\
1 & 32 & 36 & 1152 & 100000 & 782.32 & 716.45 \\
1 & 32 & 36 & 1152 & 600000 & 4630.45 & 4506.59 \\
1 & 32 & 36 & 1152 & 1000000 & 7904.56 & 7807.02 \\
      \midrule
      \multicolumn{7}{c}{Bcast, $2$-ported} \\
2 & 32 & 36 & 1152 & 1 & 22.16 & 12.64 \\
2 & 32 & 36 & 1152 & 6 & 17.57 & 12.16 \\
2 & 32 & 36 & 1152 & 10 & 15.48 & 11.68 \\
2 & 32 & 36 & 1152 & 60 & 20.34 & 17.40 \\
2 & 32 & 36 & 1152 & 100 & 21.59 & 18.12 \\
2 & 32 & 36 & 1152 & 600 & 36.39 & 25.75 \\
2 & 32 & 36 & 1152 & 1000 & 32.52 & 28.13 \\
2 & 32 & 36 & 1152 & 6000 & 125.51 & 93.46 \\
2 & 32 & 36 & 1152 & 10000 & 147.12 & 119.45 \\
2 & 32 & 36 & 1152 & 60000 & 477.82 & 459.43 \\
2 & 32 & 36 & 1152 & 100000 & 721.66 & 697.85 \\
2 & 32 & 36 & 1152 & 600000 & 4957.31 & 4784.35 \\
2 & 32 & 36 & 1152 & 1000000 & 8383.37 & 8166.79 \\
      \midrule
      \multicolumn{7}{c}{Bcast, $3$-ported} \\
3 & 32 & 36 & 1152 & 1 & 26.31 & 13.83 \\
3 & 32 & 36 & 1152 & 6 & 24.07 & 11.68 \\
3 & 32 & 36 & 1152 & 10 & 17.95 & 12.40 \\
3 & 32 & 36 & 1152 & 60 & 22.63 & 15.02 \\
3 & 32 & 36 & 1152 & 100 & 20.90 & 15.50 \\
3 & 32 & 36 & 1152 & 600 & 36.23 & 22.89 \\
3 & 32 & 36 & 1152 & 1000 & 31.83 & 25.99 \\
3 & 32 & 36 & 1152 & 6000 & 107.28 & 86.31 \\
3 & 32 & 36 & 1152 & 10000 & 125.14 & 113.25 \\
3 & 32 & 36 & 1152 & 60000 & 508.11 & 494.24 \\
3 & 32 & 36 & 1152 & 100000 & 814.43 & 788.69 \\
3 & 32 & 36 & 1152 & 600000 & 5397.68 & 5127.91 \\
3 & 32 & 36 & 1152 & 1000000 & 9070.13 & 8935.21 \\
    \bottomrule
    \end{tabular}
    \end{center}
    \end{small}
\end{table}

\begin{table}
  \caption{Results for $k$-ported Bcast for $k=4,5,6$ on the
    ``Hydra'' system.  The MPI library used is \mpichversion.}
    \label{tab:bcast.port.n32.mpich.B}
    \begin{small}
    \begin{center}
    \begin{tabular}{rrrrrrr}
      \toprule
$k$ & $n$ & $N$ & $p$ & $c$ & avg ($\mu s$) & min ($\mu s$) \\
      \midrule
      \multicolumn{7}{c}{Bcast, $4$-ported} \\
4 & 32 & 36 & 1152 & 1 & 39.30 & 12.87 \\
4 & 32 & 36 & 1152 & 6 & 21.71 & 13.35 \\
4 & 32 & 36 & 1152 & 10 & 23.64 & 13.35 \\
4 & 32 & 36 & 1152 & 60 & 27.79 & 15.74 \\
4 & 32 & 36 & 1152 & 100 & 24.95 & 16.69 \\
4 & 32 & 36 & 1152 & 600 & 40.97 & 22.65 \\
4 & 32 & 36 & 1152 & 1000 & 36.18 & 25.51 \\
4 & 32 & 36 & 1152 & 6000 & 127.19 & 92.03 \\
4 & 32 & 36 & 1152 & 10000 & 138.78 & 115.63 \\
4 & 32 & 36 & 1152 & 60000 & 508.36 & 484.23 \\
4 & 32 & 36 & 1152 & 100000 & 797.54 & 776.05 \\
4 & 32 & 36 & 1152 & 600000 & 5412.22 & 5291.94 \\
4 & 32 & 36 & 1152 & 1000000 & 9489.13 & 9166.24 \\
      \midrule
      \multicolumn{7}{c}{Bcast, $5$-ported} \\
5 & 32 & 36 & 1152 & 1 & 109.23 & 13.83 \\
5 & 32 & 36 & 1152 & 6 & 19.52 & 11.44 \\
5 & 32 & 36 & 1152 & 10 & 19.42 & 11.92 \\
5 & 32 & 36 & 1152 & 60 & 16.99 & 14.31 \\
5 & 32 & 36 & 1152 & 100 & 21.94 & 14.78 \\
5 & 32 & 36 & 1152 & 600 & 33.24 & 25.03 \\
5 & 32 & 36 & 1152 & 1000 & 36.54 & 28.61 \\
5 & 32 & 36 & 1152 & 6000 & 107.40 & 89.88 \\
5 & 32 & 36 & 1152 & 10000 & 137.68 & 125.65 \\
5 & 32 & 36 & 1152 & 60000 & 628.66 & 620.84 \\
5 & 32 & 36 & 1152 & 100000 & 1010.39 & 1002.55 \\
5 & 32 & 36 & 1152 & 600000 & 6304.32 & 6212.95 \\
5 & 32 & 36 & 1152 & 1000000 & 10668.45 & 10620.83 \\
      \midrule
      \multicolumn{7}{c}{Bcast, $6$-ported} \\
6 & 32 & 36 & 1152 & 1 & 38.04 & 13.83 \\
6 & 32 & 36 & 1152 & 6 & 19.40 & 13.83 \\
6 & 32 & 36 & 1152 & 10 & 16.08 & 13.59 \\
6 & 32 & 36 & 1152 & 60 & 27.36 & 15.74 \\
6 & 32 & 36 & 1152 & 100 & 23.87 & 16.21 \\
6 & 32 & 36 & 1152 & 600 & 33.34 & 24.08 \\
6 & 32 & 36 & 1152 & 1000 & 32.08 & 28.13 \\
6 & 32 & 36 & 1152 & 6000 & 107.62 & 92.27 \\
6 & 32 & 36 & 1152 & 10000 & 136.82 & 123.02 \\
6 & 32 & 36 & 1152 & 60000 & 623.62 & 606.78 \\
6 & 32 & 36 & 1152 & 100000 & 998.41 & 982.52 \\
6 & 32 & 36 & 1152 & 600000 & 6354.42 & 6099.46 \\
6 & 32 & 36 & 1152 & 1000000 & 10785.03 & 10522.60 \\
    \bottomrule
    \end{tabular}
    \end{center}
    \end{small}
\end{table}

\begin{table}
  \caption{Results for full-lane Bcast and the native \mpibcast on the
    ``Hydra'' system.  The MPI library used is \mpichversion.}
    \label{tab:bcast.full.n32.mpich}
    \begin{small}
    \begin{center}
    \begin{tabular}{rrrrrrr}
      \toprule
$k$ & $n$ & $N$ & $p$ & $c$ & avg ($\mu s$) & min ($\mu s$) \\
      \midrule
      \multicolumn{7}{c}{Full-lane Bcast} \\
6 & 32 & 36 & 1152 & 1 & 31.17 & 26.46 \\
6 & 32 & 36 & 1152 & 6 & 29.93 & 25.99 \\
6 & 32 & 36 & 1152 & 10 & 29.49 & 25.51 \\
6 & 32 & 36 & 1152 & 60 & 56.39 & 50.31 \\
6 & 32 & 36 & 1152 & 100 & 51.48 & 42.92 \\
6 & 32 & 36 & 1152 & 600 & 57.47 & 51.50 \\
6 & 32 & 36 & 1152 & 1000 & 58.88 & 51.98 \\
6 & 32 & 36 & 1152 & 6000 & 73.84 & 66.04 \\
6 & 32 & 36 & 1152 & 10000 & 91.65 & 76.77 \\
6 & 32 & 36 & 1152 & 60000 & 251.97 & 217.44 \\
6 & 32 & 36 & 1152 & 100000 & 456.89 & 365.50 \\
6 & 32 & 36 & 1152 & 600000 & 2916.50 & 2862.69 \\
6 & 32 & 36 & 1152 & 1000000 & 4878.80 & 4687.07 \\
      \midrule
      \multicolumn{7}{c}{\mpibcast} \\
6 & 32 & 36 & 1152 & 1 & 12.79 & 7.39 \\
6 & 32 & 36 & 1152 & 6 & 12.72 & 9.54 \\
6 & 32 & 36 & 1152 & 10 & 12.65 & 9.06 \\
6 & 32 & 36 & 1152 & 60 & 20.52 & 17.64 \\
6 & 32 & 36 & 1152 & 100 & 20.18 & 18.12 \\
6 & 32 & 36 & 1152 & 600 & 30.35 & 24.32 \\
6 & 32 & 36 & 1152 & 1000 & 33.20 & 29.56 \\
6 & 32 & 36 & 1152 & 6000 & 169.00 & 161.65 \\
6 & 32 & 36 & 1152 & 10000 & 192.53 & 182.15 \\
6 & 32 & 36 & 1152 & 60000 & 408.01 & 384.09 \\
6 & 32 & 36 & 1152 & 100000 & 571.58 & 534.77 \\
6 & 32 & 36 & 1152 & 600000 & 3790.56 & 3711.22 \\
6 & 32 & 36 & 1152 & 1000000 & 5779.13 & 5672.69 \\
    \bottomrule
    \end{tabular}
    \end{center}
    \end{small}
\end{table}

\clearpage

\subsection{Scatter}

We compare the $k$-lane and the $k$-ported scatter implementations
against the full-lane implementation, and the native \mpiscatter for
the three different MPI libraries, and try with $k=1,2,3,4,5,6$
virtual lanes.  The count $c$ is the number of data elements per
process. Data elements are here \mpiint.  For the running times (in
$\mu$-seconds, measured with \mpibarrier and \mpiwtime), we report
both average and minimum time of the slowest process over \repetitions
repetitions with \warmup initial, not measured warm-up repetitions.

The scatter results with \openmpiversion are shown in
Table~\ref{tab:scatter.lane.n32}, Table~\ref{tab:scatter.lane.n32.B},
Table~\ref{tab:scatter.port.n32}, Table~\ref{tab:scatter.port.n32.B}
and Table~\ref{tab:scatter.full.n32}. The $k$-lane results are
disappointing, with slightly increasing running times with increasing
$k$. The $k$-ported algorithm fare better, with small, but significant
decreases in running times with increasing $k$. This is contradictory
to our expectations. Both algorithms are significantly better,
though, than both full-lane algorithm and \mpiscatter.

\begin{table}
  \caption{Results for $k$-lane Scatter for $k=1,2,3$ on the
    ``Hydra'' system.  The MPI library used is \openmpiversion.}
    \label{tab:scatter.lane.n32}
    \begin{small}
    \begin{center}
    \begin{tabular}{rrrrrrr}
      \toprule
$k$ & $n$ & $N$ & $p$ & $c$ & avg ($\mu s$) & min ($\mu s$) \\
      \midrule
      \multicolumn{7}{c}{Scatter, $1$ lane} \\
1 & 32 & 36 & 1152 & 1 & 17.52 & 14.58 \\
1 & 32 & 36 & 1152 & 6 & 27.35 & 23.33 \\
1 & 32 & 36 & 1152 & 9 & 32.92 & 27.98 \\
1 & 32 & 36 & 1152 & 53 & 75.92 & 67.97 \\
1 & 32 & 36 & 1152 & 87 & 131.96 & 123.82 \\
1 & 32 & 36 & 1152 & 521 & 344.98 & 334.52 \\
1 & 32 & 36 & 1152 & 869 & 458.39 & 448.11 \\
      \midrule
      \multicolumn{7}{c}{Scatter, $2$ lanes} \\
2 & 32 & 36 & 1152 & 1 & 20.21 & 17.29 \\
2 & 32 & 36 & 1152 & 6 & 28.73 & 25.13 \\
2 & 32 & 36 & 1152 & 9 & 34.92 & 31.92 \\
2 & 32 & 36 & 1152 & 53 & 69.03 & 65.18 \\
2 & 32 & 36 & 1152 & 87 & 148.96 & 113.72 \\
2 & 32 & 36 & 1152 & 521 & 405.99 & 357.42 \\
2 & 32 & 36 & 1152 & 869 & 538.72 & 533.98 \\
      \midrule
      \multicolumn{7}{c}{Scatter, $3$ lanes} \\
3 & 32 & 36 & 1152 & 1 & 18.50 & 15.55 \\
3 & 32 & 36 & 1152 & 6 & 28.21 & 23.41 \\
3 & 32 & 36 & 1152 & 9 & 32.90 & 29.64 \\
3 & 32 & 36 & 1152 & 53 & 72.88 & 69.28 \\
3 & 32 & 36 & 1152 & 87 & 134.10 & 123.58 \\
3 & 32 & 36 & 1152 & 521 & 403.53 & 391.47 \\
3 & 32 & 36 & 1152 & 869 & 614.59 & 605.06 \\
    \bottomrule
    \end{tabular}
    \end{center}
    \end{small}
\end{table}

\begin{table}
  \caption{Results for $k$-lane Scatter for $k=4,5,6$ on the
    ``Hydra'' system.  The MPI library used is \openmpiversion.}
    \label{tab:scatter.lane.n32.B}
    \begin{small}
    \begin{center}
    \begin{tabular}{rrrrrrr}
      \toprule
$k$ & $n$ & $N$ & $p$ & $c$ & avg ($\mu s$) & min ($\mu s$) \\
      \midrule
      \multicolumn{7}{c}{Scatter, $4$ lanes} \\
4 & 32 & 36 & 1152 & 1 & 19.09 & 15.55 \\
4 & 32 & 36 & 1152 & 6 & 29.11 & 25.39 \\
4 & 32 & 36 & 1152 & 9 & 33.89 & 29.65 \\
4 & 32 & 36 & 1152 & 53 & 70.54 & 66.72 \\
4 & 32 & 36 & 1152 & 87 & 125.99 & 113.33 \\
4 & 32 & 36 & 1152 & 521 & 344.56 & 326.98 \\
4 & 32 & 36 & 1152 & 869 & 503.84 & 495.55 \\
      \midrule
      \multicolumn{7}{c}{Scatter, $5$ lanes} \\
5 & 32 & 36 & 1152 & 1 & 18.09 & 15.00 \\
5 & 32 & 36 & 1152 & 6 & 29.93 & 25.36 \\
5 & 32 & 36 & 1152 & 9 & 33.54 & 28.53 \\
5 & 32 & 36 & 1152 & 53 & 74.79 & 67.45 \\
5 & 32 & 36 & 1152 & 87 & 136.05 & 124.00 \\
5 & 32 & 36 & 1152 & 521 & 371.11 & 360.75 \\
5 & 32 & 36 & 1152 & 869 & 540.45 & 525.81 \\
      \midrule
      \multicolumn{7}{c}{Scatter, $6$ lanes} \\
6 & 32 & 36 & 1152 & 1 & 18.45 & 15.81 \\
6 & 32 & 36 & 1152 & 6 & 27.53 & 23.39 \\
6 & 32 & 36 & 1152 & 9 & 32.62 & 29.33 \\
6 & 32 & 36 & 1152 & 53 & 70.53 & 63.27 \\
6 & 32 & 36 & 1152 & 87 & 116.74 & 106.34 \\
6 & 32 & 36 & 1152 & 521 & 316.66 & 307.23 \\
6 & 32 & 36 & 1152 & 869 & 460.32 & 454.83 \\
    \bottomrule
    \end{tabular}
    \end{center}
    \end{small}
\end{table}

\begin{table}
  \caption{Results for $k$-ported Scatter for $k=1,2,3$ on the
    ``Hydra'' system.  The MPI library used is \openmpiversion.}
    \label{tab:scatter.port.n32}
    \begin{small}
    \begin{center}
    \begin{tabular}{rrrrrrr}
      \toprule
$k$ & $n$ & $N$ & $p$ & $c$ & avg ($\mu s$) & min ($\mu s$) \\
      \midrule
      \multicolumn{7}{c}{Scatter, $1$-ported} \\
1 & 32 & 36 & 1152 & 1 & 27.90 & 19.41 \\
1 & 32 & 36 & 1152 & 6 & 31.55 & 27.37 \\
1 & 32 & 36 & 1152 & 9 & 37.92 & 34.91 \\
1 & 32 & 36 & 1152 & 53 & 78.08 & 71.71 \\
1 & 32 & 36 & 1152 & 87 & 99.87 & 96.51 \\
1 & 32 & 36 & 1152 & 521 & 306.76 & 295.57 \\
1 & 32 & 36 & 1152 & 869 & 453.82 & 439.72 \\
      \midrule
      \multicolumn{7}{c}{Scatter, $2$-ported} \\
2 & 32 & 36 & 1152 & 1 & 17.38 & 14.40 \\
2 & 32 & 36 & 1152 & 6 & 23.67 & 20.92 \\
2 & 32 & 36 & 1152 & 9 & 29.19 & 23.01 \\
2 & 32 & 36 & 1152 & 53 & 61.11 & 58.15 \\
2 & 32 & 36 & 1152 & 87 & 81.52 & 76.45 \\
2 & 32 & 36 & 1152 & 521 & 268.79 & 262.14 \\
2 & 32 & 36 & 1152 & 869 & 432.82 & 400.04 \\
      \midrule
      \multicolumn{7}{c}{Scatter, $3$-ported} \\
3 & 32 & 36 & 1152 & 1 & 16.15 & 13.61 \\
3 & 32 & 36 & 1152 & 6 & 20.14 & 17.77 \\
3 & 32 & 36 & 1152 & 9 & 23.46 & 21.46 \\
3 & 32 & 36 & 1152 & 53 & 52.28 & 46.03 \\
3 & 32 & 36 & 1152 & 87 & 76.68 & 70.21 \\
3 & 32 & 36 & 1152 & 521 & 256.02 & 250.10 \\
3 & 32 & 36 & 1152 & 869 & 390.82 & 381.91 \\
    \bottomrule
    \end{tabular}
    \end{center}
    \end{small}
\end{table}

\begin{table}
  \caption{Results for $k$-ported Scatter for $k=4,5,6$ on the
    ``Hydra'' system.  The MPI library used is \openmpiversion.}
    \label{tab:scatter.port.n32.B}
    \begin{small}
    \begin{center}
    \begin{tabular}{rrrrrrr}
      \toprule
$k$ & $n$ & $N$ & $p$ & $c$ & avg ($\mu s$) & min ($\mu s$) \\
      \midrule
      \multicolumn{7}{c}{Scatter, $4$-ported} \\
4 & 32 & 36 & 1152 & 1 & 14.76 & 12.74 \\
4 & 32 & 36 & 1152 & 6 & 19.12 & 16.47 \\
4 & 32 & 36 & 1152 & 9 & 21.57 & 18.61 \\
4 & 32 & 36 & 1152 & 53 & 46.52 & 43.40 \\
4 & 32 & 36 & 1152 & 87 & 74.16 & 68.93 \\
4 & 32 & 36 & 1152 & 521 & 247.63 & 243.01 \\
4 & 32 & 36 & 1152 & 869 & 390.54 & 379.67 \\
      \midrule
      \multicolumn{7}{c}{Scatter, $5$-ported} \\
5 & 32 & 36 & 1152 & 1 & 14.15 & 11.84 \\
5 & 32 & 36 & 1152 & 6 & 17.40 & 15.40 \\
5 & 32 & 36 & 1152 & 9 & 21.19 & 18.23 \\
5 & 32 & 36 & 1152 & 53 & 49.01 & 46.12 \\
5 & 32 & 36 & 1152 & 87 & 68.90 & 66.10 \\
5 & 32 & 36 & 1152 & 521 & 258.49 & 251.34 \\
5 & 32 & 36 & 1152 & 869 & 397.27 & 376.72 \\
      \midrule
      \multicolumn{7}{c}{Scatter, $6$-ported} \\
6 & 32 & 36 & 1152 & 1 & 18.05 & 13.19 \\
6 & 32 & 36 & 1152 & 6 & 18.44 & 15.70 \\
6 & 32 & 36 & 1152 & 9 & 23.81 & 18.72 \\
6 & 32 & 36 & 1152 & 53 & 48.89 & 46.26 \\
6 & 32 & 36 & 1152 & 87 & 64.08 & 58.93 \\
6 & 32 & 36 & 1152 & 521 & 249.04 & 239.69 \\
6 & 32 & 36 & 1152 & 869 & 388.39 & 380.78 \\
    \bottomrule
    \end{tabular}
    \end{center}
    \end{small}
\end{table}

\begin{table}
  \caption{Results for full-lane Scatter and the native \mpiscatter on the
    ``Hydra'' system.  The MPI library used is \openmpiversion.}
    \label{tab:scatter.full.n32}
    \begin{small}
    \begin{center}
    \begin{tabular}{rrrrrrr}
      \toprule
$k$ & $n$ & $N$ & $p$ & $c$ & avg ($\mu s$) & min ($\mu s$) \\
\midrule
\multicolumn{7}{c}{Full-lane Scatter} \\
6 & 32 & 36 & 1152 & 1 & 28.16 & 25.11 \\
6 & 32 & 36 & 1152 & 6 & 40.89 & 38.45 \\
6 & 32 & 36 & 1152 & 9 & 47.68 & 43.31 \\
6 & 32 & 36 & 1152 & 53 & 97.17 & 91.66 \\
6 & 32 & 36 & 1152 & 87 & 161.30 & 145.27 \\
6 & 32 & 36 & 1152 & 521 & 962.05 & 944.55 \\
6 & 32 & 36 & 1152 & 869 & 1444.02 & 1414.48 \\
\midrule
\multicolumn{7}{c}{\mpiscatter} \\
6 & 32 & 36 & 1152 & 1 & 26.53 & 21.42 \\
6 & 32 & 36 & 1152 & 6 & 35.49 & 31.52 \\
6 & 32 & 36 & 1152 & 9 & 46.06 & 41.83 \\
6 & 32 & 36 & 1152 & 53 & 120.14 & 114.73 \\
6 & 32 & 36 & 1152 & 87 & 483.48 & 449.04 \\
6 & 32 & 36 & 1152 & 521 & 754.45 & 746.91 \\
6 & 32 & 36 & 1152 & 869 & 1001.17 & 985.02 \\
    \bottomrule
    \end{tabular}
    \end{center}
    \end{small}
\end{table}

The scatter results with \intelmpiversion are shown in
Table~\ref{tab:scatter.lane.n32.intel},
Table~\ref{tab:scatter.lane.n32.intel.B},
Table~\ref{tab:scatter.port.n32.intel},
Table~\ref{tab:scatter.port.n32.intel.B} and
Table~\ref{tab:scatter.full.n32.intel}.
The results are comparable to the results with \openmpiversion.

\begin{table}
  \caption{Results for $k$-lane Scatter for $k=1,2,3$ on the
    ``Hydra'' system.  The MPI library used is \intelmpiversion.}
    \label{tab:scatter.lane.n32.intel}
    \begin{small}
    \begin{center}
    \begin{tabular}{rrrrrrr}
      \toprule
$k$ & $n$ & $N$ & $p$ & $c$ & avg ($\mu s$) & min ($\mu s$) \\
\midrule
\multicolumn{7}{c}{Scatter, $1$ lane} \\
1 & 32 & 36 & 1152 & 1 & 20.67 & 19.07 \\
1 & 32 & 36 & 1152 & 6 & 26.29 & 24.08 \\
1 & 32 & 36 & 1152 & 9 & 33.87 & 30.99 \\
1 & 32 & 36 & 1152 & 53 & 71.29 & 69.86 \\
1 & 32 & 36 & 1152 & 87 & 95.79 & 94.18 \\
1 & 32 & 36 & 1152 & 521 & 310.16 & 307.08 \\
1 & 32 & 36 & 1152 & 869 & 448.29 & 444.17 \\
\midrule
\multicolumn{7}{c}{Scatter, $2$ lanes} \\
2 & 32 & 36 & 1152 & 1 & 18.93 & 15.97 \\
2 & 32 & 36 & 1152 & 6 & 24.95 & 22.17 \\
2 & 32 & 36 & 1152 & 9 & 28.06 & 25.99 \\
2 & 32 & 36 & 1152 & 53 & 72.75 & 70.10 \\
2 & 32 & 36 & 1152 & 87 & 97.97 & 93.94 \\
2 & 32 & 36 & 1152 & 521 & 350.63 & 341.89 \\
2 & 32 & 36 & 1152 & 869 & 544.52 & 535.96 \\
\midrule
\multicolumn{7}{c}{Scatter, $3$ lanes} \\
3 & 32 & 36 & 1152 & 1 & 16.79 & 15.02 \\
3 & 32 & 36 & 1152 & 6 & 22.01 & 19.07 \\
3 & 32 & 36 & 1152 & 9 & 24.74 & 22.17 \\
3 & 32 & 36 & 1152 & 53 & 63.82 & 61.04 \\
3 & 32 & 36 & 1152 & 87 & 91.46 & 87.02 \\
3 & 32 & 36 & 1152 & 521 & 375.12 & 331.16 \\
3 & 32 & 36 & 1152 & 869 & 606.53 & 588.89 \\
    \bottomrule
    \end{tabular}
    \end{center}
    \end{small}
\end{table}

\begin{table}
  \caption{Results for $k$-lane Scatter for $k=4,5,6$ on the
    ``Hydra'' system.  The MPI library used is \intelmpiversion.}
    \label{tab:scatter.lane.n32.intel.B}
    \begin{small}
    \begin{center}
    \begin{tabular}{rrrrrrr}
      \toprule
$k$ & $n$ & $N$ & $p$ & $c$ & avg ($\mu s$) & min ($\mu s$) \\
      \midrule
      \multicolumn{7}{c}{Scatter, $4$ lanes} \\
4 & 32 & 36 & 1152 & 1 & 18.62 & 14.07 \\
4 & 32 & 36 & 1152 & 6 & 22.71 & 19.07 \\
4 & 32 & 36 & 1152 & 9 & 24.11 & 21.93 \\
4 & 32 & 36 & 1152 & 53 & 65.73 & 61.99 \\
4 & 32 & 36 & 1152 & 87 & 102.46 & 93.94 \\
4 & 32 & 36 & 1152 & 521 & 414.09 & 397.92 \\
4 & 32 & 36 & 1152 & 869 & 659.47 & 645.88 \\
      \midrule
      \multicolumn{7}{c}{Scatter, $5$ lanes} \\
5 & 32 & 36 & 1152 & 1 & 15.41 & 13.11 \\
5 & 32 & 36 & 1152 & 6 & 19.37 & 16.93 \\
5 & 32 & 36 & 1152 & 9 & 22.39 & 19.07 \\
5 & 32 & 36 & 1152 & 53 & 57.17 & 54.12 \\
5 & 32 & 36 & 1152 & 87 & 93.04 & 87.98 \\
5 & 32 & 36 & 1152 & 521 & 401.56 & 378.85 \\
5 & 32 & 36 & 1152 & 869 & 632.98 & 603.91 \\
      \midrule
      \multicolumn{7}{c}{Scatter, $6$ lanes} \\
6 & 32 & 36 & 1152 & 1 & 14.93 & 12.16 \\
6 & 32 & 36 & 1152 & 6 & 19.29 & 16.93 \\
6 & 32 & 36 & 1152 & 9 & 21.70 & 19.07 \\
6 & 32 & 36 & 1152 & 53 & 58.26 & 55.07 \\
6 & 32 & 36 & 1152 & 87 & 84.60 & 81.06 \\
6 & 32 & 36 & 1152 & 521 & 379.50 & 353.10 \\
6 & 32 & 36 & 1152 & 869 & 590.42 & 556.95 \\
    \bottomrule
    \end{tabular}
    \end{center}
    \end{small}
\end{table}

\begin{table}
  \caption{Results for $k$-ported Scatter for $k=1,2,3$ on the
    ``Hydra'' system.  The MPI library used is \intelmpiversion.}
    \label{tab:scatter.port.n32.intel}
    \begin{small}
    \begin{center}
    \begin{tabular}{rrrrrrr}
      \toprule
$k$ & $n$ & $N$ & $p$ & $c$ & avg ($\mu s$) & min ($\mu s$) \\
      \midrule
      \multicolumn{7}{c}{Scatter, $1$-ported} \\
1 & 32 & 36 & 1152 & 1 & 23.90 & 19.07 \\
1 & 32 & 36 & 1152 & 6 & 29.18 & 26.94 \\
1 & 32 & 36 & 1152 & 9 & 38.74 & 33.86 \\
1 & 32 & 36 & 1152 & 53 & 74.39 & 71.05 \\
1 & 32 & 36 & 1152 & 87 & 100.49 & 95.13 \\
1 & 32 & 36 & 1152 & 521 & 298.80 & 293.97 \\
1 & 32 & 36 & 1152 & 869 & 464.12 & 437.97 \\
      \midrule
      \multicolumn{7}{c}{Scatter, $2$-ported} \\
2 & 32 & 36 & 1152 & 1 & 15.34 & 13.83 \\
2 & 32 & 36 & 1152 & 6 & 22.85 & 19.79 \\
2 & 32 & 36 & 1152 & 9 & 24.91 & 20.98 \\
2 & 32 & 36 & 1152 & 53 & 57.52 & 56.03 \\
2 & 32 & 36 & 1152 & 87 & 75.21 & 72.96 \\
2 & 32 & 36 & 1152 & 521 & 273.80 & 267.03 \\
2 & 32 & 36 & 1152 & 869 & 420.82 & 414.13 \\
      \midrule
      \multicolumn{7}{c}{Scatter, $3$-ported} \\
3 & 32 & 36 & 1152 & 1 & 14.33 & 12.87 \\
3 & 32 & 36 & 1152 & 6 & 18.12 & 16.93 \\
3 & 32 & 36 & 1152 & 9 & 20.93 & 19.07 \\
3 & 32 & 36 & 1152 & 53 & 62.84 & 43.87 \\
3 & 32 & 36 & 1152 & 87 & 123.04 & 67.95 \\
3 & 32 & 36 & 1152 & 521 & 285.47 & 248.91 \\
3 & 32 & 36 & 1152 & 869 & 403.15 & 386.95 \\
    \bottomrule
    \end{tabular}
    \end{center}
    \end{small}
\end{table}

\begin{table}
  \caption{Results for $k$-ported Scatter for $k=4,5,6$ on the
    ``Hydra'' system.  The MPI library used is \intelmpiversion.}
    \label{tab:scatter.port.n32.intel.B}
    \begin{small}
    \begin{center}
    \begin{tabular}{rrrrrrr}
      \toprule
$k$ & $n$ & $N$ & $p$ & $c$ & avg ($\mu s$) & min ($\mu s$) \\
\midrule
\multicolumn{7}{c}{Scatter, $1$-ported} \\
4 & 32 & 36 & 1152 & 1 & 13.62 & 11.92 \\
4 & 32 & 36 & 1152 & 6 & 17.75 & 15.97 \\
4 & 32 & 36 & 1152 & 9 & 19.97 & 17.88 \\
4 & 32 & 36 & 1152 & 53 & 43.88 & 41.96 \\
4 & 32 & 36 & 1152 & 87 & 68.20 & 66.04 \\
4 & 32 & 36 & 1152 & 521 & 249.33 & 245.09 \\
4 & 32 & 36 & 1152 & 869 & 390.04 & 386.00 \\
\midrule
\multicolumn{7}{c}{Scatter, $2$-ported} \\
5 & 32 & 36 & 1152 & 1 & 12.70 & 10.01 \\
5 & 32 & 36 & 1152 & 6 & 15.93 & 13.11 \\
5 & 32 & 36 & 1152 & 9 & 17.66 & 15.97 \\
5 & 32 & 36 & 1152 & 53 & 44.23 & 42.20 \\
5 & 32 & 36 & 1152 & 87 & 64.80 & 62.94 \\
5 & 32 & 36 & 1152 & 521 & 248.61 & 241.99 \\
5 & 32 & 36 & 1152 & 869 & 416.05 & 376.94 \\
\midrule
\multicolumn{7}{c}{Scatter, $3$-ported} \\
6 & 32 & 36 & 1152 & 1 & 15.25 & 11.92 \\
6 & 32 & 36 & 1152 & 6 & 17.96 & 15.02 \\
6 & 32 & 36 & 1152 & 9 & 18.40 & 15.97 \\
6 & 32 & 36 & 1152 & 53 & 48.56 & 42.92 \\
6 & 32 & 36 & 1152 & 87 & 56.09 & 53.17 \\
6 & 32 & 36 & 1152 & 521 & 237.90 & 235.08 \\
6 & 32 & 36 & 1152 & 869 & 377.11 & 375.03 \\
    \bottomrule
    \end{tabular}
    \end{center}
    \end{small}
\end{table}

\begin{table}
  \caption{Results for full-lane Scatter and the native \mpiscatter on the
    ``Hydra'' system.  The MPI library used is \intelmpiversion.}
    \label{tab:scatter.full.n32.intel}
    \begin{small}
    \begin{center}
    \begin{tabular}{rrrrrrr}
      \toprule
$k$ & $n$ & $N$ & $p$ & $c$ & avg ($\mu s$) & min ($\mu s$) \\
      \midrule
      \multicolumn{7}{c}{Full-lane Scatter} \\
6 & 32 & 36 & 1152 & 1 & 17.32 & 14.07 \\
6 & 32 & 36 & 1152 & 6 & 28.67 & 26.94 \\
6 & 32 & 36 & 1152 & 9 & 52.53 & 50.78 \\
6 & 32 & 36 & 1152 & 53 & 104.50 & 102.04 \\
6 & 32 & 36 & 1152 & 87 & 148.01 & 144.00 \\
6 & 32 & 36 & 1152 & 521 & 550.44 & 540.02 \\
6 & 32 & 36 & 1152 & 869 & 789.97 & 773.19 \\
      \midrule
      \multicolumn{7}{c}{\mpiscatter} \\
6 & 32 & 36 & 1152 & 1 & 18.63 & 16.93 \\
6 & 32 & 36 & 1152 & 6 & 25.10 & 22.89 \\
6 & 32 & 36 & 1152 & 9 & 30.11 & 28.13 \\
6 & 32 & 36 & 1152 & 53 & 538.26 & 530.00 \\
6 & 32 & 36 & 1152 & 87 & 552.78 & 545.98 \\
6 & 32 & 36 & 1152 & 521 & 750.22 & 741.96 \\
6 & 32 & 36 & 1152 & 869 & 890.13 & 879.05 \\
    \bottomrule
    \end{tabular}
    \end{center}
    \end{small}
\end{table}

The scatter results with \mpichversion are shown in
Table~\ref{tab:scatter.lane.n32.mpich},
Table~\ref{tab:scatter.lane.n32.mpich.B},
Table~\ref{tab:scatter.port.n32.mpich},
Table~\ref{tab:scatter.port.n32.mpich.B} and
Table~\ref{tab:scatter.full.n32.mpich}.  The results are comparable to
the results with \openmpiversion and \intelmpiversion.

\begin{table}
  \caption{Results for $k$-lane Scatter for $k=1,2,3$ on the
    ``Hydra'' system.  The MPI library used is \mpichversion.}
    \label{tab:scatter.lane.n32.mpich}
    \begin{small}
    \begin{center}
    \begin{tabular}{rrrrrrr}
      \toprule
$k$ & $n$ & $N$ & $p$ & $c$ & avg ($\mu s$) & min ($\mu s$) \\
      \midrule
      \multicolumn{7}{c}{Scatter, $1$ lane} \\
1 & 32 & 36 & 1152 & 1 & 21.69 & 19.79 \\
1 & 32 & 36 & 1152 & 6 & 27.27 & 24.80 \\
1 & 32 & 36 & 1152 & 9 & 33.30 & 31.47 \\
1 & 32 & 36 & 1152 & 53 & 72.32 & 69.62 \\
1 & 32 & 36 & 1152 & 87 & 96.53 & 94.18 \\
1 & 32 & 36 & 1152 & 521 & 310.95 & 305.18 \\
1 & 32 & 36 & 1152 & 869 & 451.11 & 443.22 \\
      \midrule
      \multicolumn{7}{c}{Scatter, $2$ lanes} \\
2 & 32 & 36 & 1152 & 1 & 28.38 & 18.84 \\
2 & 32 & 36 & 1152 & 6 & 30.23 & 24.80 \\
2 & 32 & 36 & 1152 & 9 & 30.99 & 28.13 \\
2 & 32 & 36 & 1152 & 53 & 86.10 & 78.44 \\
2 & 32 & 36 & 1152 & 87 & 113.12 & 106.81 \\
2 & 32 & 36 & 1152 & 521 & 449.61 & 427.01 \\
2 & 32 & 36 & 1152 & 869 & 670.55 & 652.07 \\
      \midrule
      \multicolumn{7}{c}{Scatter, $3$ lanes} \\
3 & 32 & 36 & 1152 & 1 & 20.97 & 17.40 \\
3 & 32 & 36 & 1152 & 6 & 26.19 & 22.65 \\
3 & 32 & 36 & 1152 & 9 & 30.81 & 27.42 \\
3 & 32 & 36 & 1152 & 53 & 68.78 & 60.56 \\
3 & 32 & 36 & 1152 & 87 & 102.10 & 95.37 \\
3 & 32 & 36 & 1152 & 521 & 458.11 & 444.41 \\
3 & 32 & 36 & 1152 & 869 & 728.15 & 700.00 \\
    \bottomrule
    \end{tabular}
    \end{center}
    \end{small}
\end{table}

\begin{table}
  \caption{Results for $k$-lane Scatter for $k=4,5,6$ on the
    ``Hydra'' system.  The MPI library used is \mpichversion.}
    \label{tab:scatter.lane.n32.mpich.B}
    \begin{small}
    \begin{center}
    \begin{tabular}{rrrrrrr}
      \toprule
$k$ & $n$ & $N$ & $p$ & $c$ & avg ($\mu s$) & min ($\mu s$) \\
\midrule
\multicolumn{7}{c}{Scatter, $4$ lanes} \\
4 & 32 & 36 & 1152 & 1 & 20.90 & 17.17 \\
4 & 32 & 36 & 1152 & 6 & 48.71 & 21.46 \\
4 & 32 & 36 & 1152 & 9 & 44.77 & 24.80 \\
4 & 32 & 36 & 1152 & 53 & 84.86 & 62.23 \\
4 & 32 & 36 & 1152 & 87 & 117.60 & 100.37 \\
4 & 32 & 36 & 1152 & 521 & 538.36 & 443.94 \\
4 & 32 & 36 & 1152 & 869 & 817.65 & 711.92 \\
\midrule
\multicolumn{7}{c}{Scatter, $5$ lanes} \\
5 & 32 & 36 & 1152 & 1 & 19.68 & 16.69 \\
5 & 32 & 36 & 1152 & 6 & 35.91 & 20.50 \\
5 & 32 & 36 & 1152 & 9 & 35.76 & 23.84 \\
5 & 32 & 36 & 1152 & 53 & 73.07 & 58.65 \\
5 & 32 & 36 & 1152 & 87 & 134.42 & 83.45 \\
5 & 32 & 36 & 1152 & 521 & 489.60 & 425.34 \\
5 & 32 & 36 & 1152 & 869 & 790.56 & 686.65 \\
\midrule
\multicolumn{7}{c}{Scatter, $6$ lanes} \\
6 & 32 & 36 & 1152 & 1 & 41.92 & 17.17 \\
6 & 32 & 36 & 1152 & 6 & 29.90 & 22.17 \\
6 & 32 & 36 & 1152 & 9 & 40.46 & 23.60 \\
6 & 32 & 36 & 1152 & 53 & 104.82 & 61.27 \\
6 & 32 & 36 & 1152 & 87 & 109.34 & 89.41 \\
6 & 32 & 36 & 1152 & 521 & 479.42 & 417.47 \\
6 & 32 & 36 & 1152 & 869 & 785.39 & 680.69 \\
    \bottomrule
    \end{tabular}
    \end{center}
    \end{small}
\end{table}

\begin{table}
  \caption{Results for $k$-ported Scatter for $k=1,2,3$ on the
    ``Hydra'' system.  The MPI library used is \mpichversion.}
    \label{tab:scatter.port.n32.mpich}
    \begin{small}
    \begin{center}
    \begin{tabular}{rrrrrrr}
      \toprule
$k$ & $n$ & $N$ & $p$ & $c$ & avg ($\mu s$) & min ($\mu s$) \\
\midrule
\multicolumn{7}{c}{Scatter, $1$-ported} \\
1 & 32 & 36 & 1152 & 1 & 33.43 & 23.37 \\
1 & 32 & 36 & 1152 & 6 & 53.42 & 29.09 \\
1 & 32 & 36 & 1152 & 9 & 158.78 & 37.19 \\
1 & 32 & 36 & 1152 & 53 & 198.85 & 76.53 \\
1 & 32 & 36 & 1152 & 87 & 168.27 & 99.42 \\
1 & 32 & 36 & 1152 & 521 & 371.20 & 297.07 \\
1 & 32 & 36 & 1152 & 869 & 472.15 & 441.31 \\
\midrule
\multicolumn{7}{c}{Scatter, $2$-ported} \\
2 & 32 & 36 & 1152 & 1 & 20.28 & 16.45 \\
2 & 32 & 36 & 1152 & 6 & 25.42 & 21.93 \\
2 & 32 & 36 & 1152 & 9 & 26.84 & 23.84 \\
2 & 32 & 36 & 1152 & 53 & 64.00 & 59.13 \\
2 & 32 & 36 & 1152 & 87 & 79.48 & 74.63 \\
2 & 32 & 36 & 1152 & 521 & 272.26 & 267.98 \\
2 & 32 & 36 & 1152 & 869 & 422.20 & 415.80 \\
\midrule
\multicolumn{7}{c}{Scatter, $3$-ported} \\
3 & 32 & 36 & 1152 & 1 & 19.06 & 16.69 \\
3 & 32 & 36 & 1152 & 6 & 22.90 & 20.03 \\
3 & 32 & 36 & 1152 & 9 & 26.50 & 21.93 \\
3 & 32 & 36 & 1152 & 53 & 48.96 & 46.25 \\
3 & 32 & 36 & 1152 & 87 & 72.72 & 69.14 \\
3 & 32 & 36 & 1152 & 521 & 256.56 & 248.91 \\
3 & 32 & 36 & 1152 & 869 & 392.12 & 387.91 \\
    \bottomrule
    \end{tabular}
    \end{center}
    \end{small}
\end{table}

\begin{table}
  \caption{Results for $k$-ported Scatter for $k=4,5,6$ on the
    ``Hydra'' system.  The MPI library used is \mpichversion.}
    \label{tab:scatter.port.n32.mpich.B}
    \begin{small}
    \begin{center}
    \begin{tabular}{rrrrrrr}
      \toprule
$k$ & $n$ & $N$ & $p$ & $c$ & avg ($\mu s$) & min ($\mu s$) \\
      \midrule
      \multicolumn{7}{c}{Scatter, $4$-ported} \\
4 & 32 & 36 & 1152 & 1 & 17.78 & 15.50 \\
4 & 32 & 36 & 1152 & 6 & 21.30 & 18.36 \\
4 & 32 & 36 & 1152 & 9 & 22.46 & 20.50 \\
4 & 32 & 36 & 1152 & 53 & 46.72 & 44.11 \\
4 & 32 & 36 & 1152 & 87 & 72.26 & 68.43 \\
4 & 32 & 36 & 1152 & 521 & 252.09 & 246.52 \\
4 & 32 & 36 & 1152 & 869 & 394.79 & 386.71 \\
      \midrule
      \multicolumn{7}{c}{Scatter, $5$-ported} \\
5 & 32 & 36 & 1152 & 1 & 17.30 & 15.26 \\
5 & 32 & 36 & 1152 & 6 & 19.56 & 17.64 \\
5 & 32 & 36 & 1152 & 9 & 22.42 & 19.55 \\
5 & 32 & 36 & 1152 & 53 & 48.77 & 46.25 \\
5 & 32 & 36 & 1152 & 87 & 67.17 & 64.85 \\
5 & 32 & 36 & 1152 & 521 & 250.48 & 246.76 \\
5 & 32 & 36 & 1152 & 869 & 386.88 & 382.90 \\
      \midrule
      \multicolumn{7}{c}{Scatter, $6$-ported} \\
6 & 32 & 36 & 1152 & 1 & 18.41 & 15.97 \\
6 & 32 & 36 & 1152 & 6 & 20.25 & 18.36 \\
6 & 32 & 36 & 1152 & 9 & 21.92 & 19.31 \\
6 & 32 & 36 & 1152 & 53 & 47.37 & 45.30 \\
6 & 32 & 36 & 1152 & 87 & 60.34 & 57.94 \\
6 & 32 & 36 & 1152 & 521 & 239.31 & 234.37 \\
6 & 32 & 36 & 1152 & 869 & 385.88 & 381.23 \\
    \bottomrule
    \end{tabular}
    \end{center}
    \end{small}
\end{table}

\begin{table}
  \caption{Results for full-lane Scatter and the native \mpiscatter on the
    ``Hydra'' system.  The MPI library used is \mpichversion.}
    \label{tab:scatter.full.n32.mpich}
    \begin{small}
    \begin{center}
    \begin{tabular}{rrrrrrr}
      \toprule
$k$ & $n$ & $N$ & $p$ & $c$ & avg ($\mu s$) & min ($\mu s$) \\
      \midrule
      \multicolumn{7}{c}{Full-lane Scatter} \\
6 & 32 & 36 & 1152 & 1 & 30.35 & 21.93 \\
6 & 32 & 36 & 1152 & 6 & 43.59 & 36.24 \\
6 & 32 & 36 & 1152 & 9 & 54.74 & 39.58 \\
6 & 32 & 36 & 1152 & 53 & 112.95 & 72.48 \\
6 & 32 & 36 & 1152 & 87 & 147.75 & 104.90 \\
6 & 32 & 36 & 1152 & 521 & 696.25 & 511.41 \\
6 & 32 & 36 & 1152 & 869 & 885.94 & 826.84 \\
      \midrule
      \multicolumn{7}{c}{\mpiscatter} \\
6 & 32 & 36 & 1152 & 1 & 20.81 & 17.64 \\
6 & 32 & 36 & 1152 & 6 & 27.28 & 23.60 \\
6 & 32 & 36 & 1152 & 9 & 32.84 & 29.80 \\
6 & 32 & 36 & 1152 & 53 & 77.03 & 69.62 \\
6 & 32 & 36 & 1152 & 87 & 101.15 & 96.08 \\
6 & 32 & 36 & 1152 & 521 & 305.16 & 301.84 \\
6 & 32 & 36 & 1152 & 869 & 443.14 & 438.69 \\
    \bottomrule
    \end{tabular}
    \end{center}
    \end{small}
\end{table}

\clearpage

\subsection{Alltoall}

We compare the $k$-lane with $k=32$ and the $k$-ported alltoall with
$k=1,2,3,4,5,6$ implementations against the full-lane implementation,
and the native \mpialltoall for the three different MPI libraries.
The count $c$ is the number of data elements per process. Data
elements are here \mpiint.  For the running times (in $\mu$-seconds,
measured with \mpibarrier and \mpiwtime), we report both average and
minimum time of the slowest process over \repetitions repetitions with
\warmup initial, not measured warm-up repetitions.

In the $k$-lane algorithm where all processors on a node send to all
processors on another ``next'' node and receives from all processors
on a different ``previous'' node, $k$ is not a parameter in the
implementation. There is therefore only one experiment in the tables
for the $k$-lane algorithms.

The alltoall results with \openmpiversion are shown in
Table~\ref{tab:alltoall.lane.n32}, Table~\ref{tab:alltoall.port.n32},
Table~\ref{tab:alltoall.port.n32.B} and
Table~\ref{tab:alltoall.full.n32}. The $k$-lane algorithm is always
significantly better than the $k$-ported algorithm. The results with
the $k$-ported algorithm shows significantly decreasing running times
with increasing $k$, which simply means a larger number of concurrent,
non-blocking send and receive operations. The results clearly show
that more non-blocking send-receive operations (up to some limit) is
beneficial to avoid delays. The number of non-blocking operations is
bounded by the small constant $k$. The best algorithm for small
problem sizes is the full-lane algorithm, which is also significantly
better than \mpialltoall.

\begin{table}
  \caption{Results for $k$-lane Alltoall for $k=32$ on the
    ``Hydra'' system.  The MPI library used is \openmpiversion.}
    \label{tab:alltoall.lane.n32}
    \begin{small}
    \begin{center}
    \begin{tabular}{rrrrrrr}
      \toprule
      $k$ & $n$ & $N$ & $p$ & $c$ & avg ($\mu s$) & min ($\mu s$) \\
      \midrule
    \multicolumn{7}{c}{Alltoall, $32$ virtual lanes} \\
1 & 32 & 36 & 1152 & 1 & 827.90 & 762.23 \\
1 & 32 & 36 & 1152 & 6 & 1411.98 & 1211.28 \\
1 & 32 & 36 & 1152 & 9 & 1459.78 & 1197.84 \\
1 & 32 & 36 & 1152 & 53 & 1744.20 & 1483.96 \\
1 & 32 & 36 & 1152 & 87 & 1794.75 & 1675.76 \\
1 & 32 & 36 & 1152 & 521 & 7355.07 & 7078.85 \\
1 & 32 & 36 & 1152 & 869 & 11848.12 & 11298.66 \\
    \bottomrule
    \end{tabular}
    \end{center}
    \end{small}
\end{table}

\begin{table}
  \caption{Results for $k$-ported Alltoall for $k=1,2,3$ on the
    ``Hydra'' system.  The MPI library used is \openmpiversion.}
    \label{tab:alltoall.port.n32}
    \begin{small}
    \begin{center}
    \begin{tabular}{rrrrrrr}
      \toprule
$k$ & $n$ & $N$ & $p$ & $c$ & avg ($\mu s$) & min ($\mu s$) \\
      \midrule
      \multicolumn{7}{c}{Alltoall, $1$-ported} \\
1 & 32 & 36 & 1152 & 1 & 2210.90 & 1944.20 \\
1 & 32 & 36 & 1152 & 6 & 3622.43 & 3175.22 \\
1 & 32 & 36 & 1152 & 9 & 4013.18 & 3254.45 \\
1 & 32 & 36 & 1152 & 53 & 4685.55 & 3664.99 \\
1 & 32 & 36 & 1152 & 87 & 4558.26 & 3839.11 \\
1 & 32 & 36 & 1152 & 521 & 9007.87 & 7475.03 \\
1 & 32 & 36 & 1152 & 869 & 11784.61 & 10969.10 \\
      \midrule
      \multicolumn{7}{c}{Alltoall, $2$-ported} \\
2 & 32 & 36 & 1152 & 1 & 1855.72 & 1548.44 \\
2 & 32 & 36 & 1152 & 6 & 2544.83 & 2013.37 \\
2 & 32 & 36 & 1152 & 9 & 2295.23 & 2066.66 \\
2 & 32 & 36 & 1152 & 53 & 2839.53 & 2356.16 \\
2 & 32 & 36 & 1152 & 87 & 3035.83 & 2621.88 \\
2 & 32 & 36 & 1152 & 521 & 6853.61 & 6545.05 \\
2 & 32 & 36 & 1152 & 869 & 11036.59 & 10557.39 \\
      \midrule
      \multicolumn{7}{c}{Alltoall, $3$-ported} \\
3 & 32 & 36 & 1152 & 1 & 1816.34 & 1356.17 \\
3 & 32 & 36 & 1152 & 6 & 2139.99 & 1683.69 \\
3 & 32 & 36 & 1152 & 9 & 2121.48 & 1675.49 \\
3 & 32 & 36 & 1152 & 53 & 3861.69 & 2005.88 \\
3 & 32 & 36 & 1152 & 87 & 2866.76 & 2238.85 \\
3 & 32 & 36 & 1152 & 521 & 6960.71 & 6539.43 \\
3 & 32 & 36 & 1152 & 869 & 11236.65 & 10561.70 \\
    \bottomrule
    \end{tabular}
    \end{center}
    \end{small}
\end{table}

\begin{table}
  \caption{Results for $k$-ported Alltoall for $k=4,5,6$ on the
    ``Hydra'' system.  The MPI library used is \openmpiversion.}
    \label{tab:alltoall.port.n32.B}
    \begin{small}
    \begin{center}
    \begin{tabular}{rrrrrrr}
      \toprule
$k$ & $n$ & $N$ & $p$ & $c$ & avg ($\mu s$) & min ($\mu s$) \\
      \midrule
      \multicolumn{7}{c}{Alltoall, $4$-ported} \\
4 & 32 & 36 & 1152 & 1 & 1391.85 & 1174.40 \\
4 & 32 & 36 & 1152 & 6 & 1859.69 & 1501.22 \\
4 & 32 & 36 & 1152 & 9 & 1803.34 & 1535.29 \\
4 & 32 & 36 & 1152 & 53 & 2292.54 & 1827.61 \\
4 & 32 & 36 & 1152 & 87 & 2543.57 & 2077.40 \\
4 & 32 & 36 & 1152 & 521 & 6747.86 & 6505.85 \\
4 & 32 & 36 & 1152 & 869 & 10825.52 & 10519.33 \\
      \midrule
      \multicolumn{7}{c}{Alltoall, $5$-ported} \\
5 & 32 & 36 & 1152 & 1 & 1337.51 & 992.30 \\
5 & 32 & 36 & 1152 & 6 & 1664.14 & 1424.12 \\
5 & 32 & 36 & 1152 & 9 & 1770.30 & 1455.54 \\
5 & 32 & 36 & 1152 & 53 & 2221.32 & 1745.16 \\
5 & 32 & 36 & 1152 & 87 & 2559.44 & 1946.62 \\
5 & 32 & 36 & 1152 & 521 & 7164.13 & 6517.17 \\
5 & 32 & 36 & 1152 & 869 & 11096.12 & 10539.55 \\
      \midrule
      \multicolumn{7}{c}{Alltoall, $6$-ported} \\
6 & 32 & 36 & 1152 & 1 & 1250.47 & 919.44 \\
6 & 32 & 36 & 1152 & 6 & 1635.61 & 1378.74 \\
6 & 32 & 36 & 1152 & 9 & 2978.66 & 1449.22 \\
6 & 32 & 36 & 1152 & 53 & 2082.91 & 1688.78 \\
6 & 32 & 36 & 1152 & 87 & 2249.74 & 1963.39 \\
6 & 32 & 36 & 1152 & 521 & 6760.98 & 6506.58 \\
6 & 32 & 36 & 1152 & 869 & 11187.27 & 10534.57 \\
    \bottomrule
    \end{tabular}
    \end{center}
    \end{small}
\end{table}

\begin{table}
  \caption{Results for full-lane Alltoall and the native \mpialltoall on the
    ``Hydra'' system.  The MPI library used is \openmpiversion.}
    \label{tab:alltoall.full.n32}
    \begin{small}
    \begin{center}
    \begin{tabular}{rrrrrrr}
      \toprule
$k$ & $n$ & $N$ & $p$ & $c$ & avg ($\mu s$) & min ($\mu s$) \\
      \midrule
      \multicolumn{7}{c}{Full-lane Alltoall} \\
6 & 32 & 36 & 1152 & 1 & 121.41 & 107.46 \\
6 & 32 & 36 & 1152 & 6 & 195.57 & 163.16 \\
6 & 32 & 36 & 1152 & 9 & 245.97 & 197.06 \\
6 & 32 & 36 & 1152 & 53 & 989.75 & 741.10 \\
6 & 32 & 36 & 1152 & 87 & 1564.89 & 1378.93 \\
6 & 32 & 36 & 1152 & 521 & 7217.60 & 6849.80 \\
6 & 32 & 36 & 1152 & 869 & 12233.77 & 11343.61 \\
      \midrule
      \multicolumn{7}{c}{\mpialltoall} \\
6 & 32 & 36 & 1152 & 1 & 198.32 & 185.31 \\
6 & 32 & 36 & 1152 & 6 & 391.14 & 363.42 \\
6 & 32 & 36 & 1152 & 9 & 458.24 & 443.65 \\
6 & 32 & 36 & 1152 & 53 & 75706.97 & 3288.51 \\
6 & 32 & 36 & 1152 & 87 & 83676.25 & 3556.35 \\
6 & 32 & 36 & 1152 & 521 & 166279.34 & 164829.82 \\
6 & 32 & 36 & 1152 & 869 & 12544.11 & 10535.47 \\
    \bottomrule
    \end{tabular}
    \end{center}
    \end{small}
\end{table}

The alltoall results with \intelmpiversion are shown in
Table~\ref{tab:alltoall.lane.n32.intel},
Table~\ref{tab:alltoall.port.n32.intel},
Table~\ref{tab:alltoall.port.n32.intel.B} and
Table~\ref{tab:alltoall.full.n32.intel}. The results are similar to
the results with \openmpiversion.

\begin{table}
  \caption{Results for $k$-lane Alltoall for $k=32$ on the
    ``Hydra'' system.  The MPI library used is \intelmpiversion.}
    \label{tab:alltoall.lane.n32.intel}
    \begin{small}
    \begin{center}
    \begin{tabular}{rrrrrrr}
      \toprule
$k$ & $n$ & $N$ & $p$ & $c$ & avg ($\mu s$) & min ($\mu s$) \\
      \midrule
      \multicolumn{7}{c}{Alltoall, $32$ virtual lanes} \\
1 & 32 & 36 & 1152 & 1 & 998.00 & 945.09 \\
1 & 32 & 36 & 1152 & 6 & 1297.45 & 1228.81 \\
1 & 32 & 36 & 1152 & 9 & 1273.16 & 1204.97 \\
1 & 32 & 36 & 1152 & 53 & 1762.75 & 1633.17 \\
1 & 32 & 36 & 1152 & 87 & 1855.02 & 1703.98 \\
1 & 32 & 36 & 1152 & 521 & 7250.89 & 6971.84 \\
1 & 32 & 36 & 1152 & 869 & 11707.94 & 11193.04 \\
    \bottomrule
    \end{tabular}
    \end{center}
    \end{small}
\end{table}

\begin{table}
  \caption{Results for $k$-ported Alltoall for $k=1,2,3$ on the
    ``Hydra'' system.  The MPI library used is \intelmpiversion.}
    \label{tab:alltoall.port.n32.intel}
    \begin{small}
    \begin{center}
    \begin{tabular}{rrrrrrr}
      \toprule
$k$ & $n$ & $N$ & $p$ & $c$ & avg ($\mu s$) & min ($\mu s$) \\
      \midrule
      \multicolumn{7}{c}{Alltoall, $1$-ported} \\
1 & 32 & 36 & 1152 & 1 & 2537.28 & 2297.88 \\
1 & 32 & 36 & 1152 & 6 & 2949.61 & 2618.07 \\
1 & 32 & 36 & 1152 & 9 & 3018.64 & 2653.12 \\
1 & 32 & 36 & 1152 & 53 & 4174.04 & 3808.98 \\
1 & 32 & 36 & 1152 & 87 & 4297.19 & 4038.81 \\
1 & 32 & 36 & 1152 & 521 & 7985.24 & 7452.01 \\
1 & 32 & 36 & 1152 & 869 & 11854.34 & 10892.15 \\
      \midrule
      \multicolumn{7}{c}{Alltoall, $2$-ported} \\
2 & 32 & 36 & 1152 & 1 & 2189.67 & 1975.06 \\
2 & 32 & 36 & 1152 & 6 & 2537.16 & 2333.88 \\
2 & 32 & 36 & 1152 & 9 & 2532.62 & 2389.19 \\
2 & 32 & 36 & 1152 & 53 & 2728.99 & 2539.87 \\
2 & 32 & 36 & 1152 & 87 & 2925.75 & 2691.98 \\
2 & 32 & 36 & 1152 & 521 & 7913.19 & 6539.11 \\
2 & 32 & 36 & 1152 & 869 & 11588.39 & 10568.14 \\
      \midrule
      \multicolumn{7}{c}{Alltoall, $3$-ported} \\
3 & 32 & 36 & 1152 & 1 & 2074.87 & 1796.96 \\
3 & 32 & 36 & 1152 & 6 & 2508.95 & 2108.10 \\
3 & 32 & 36 & 1152 & 9 & 2455.79 & 2161.03 \\
3 & 32 & 36 & 1152 & 53 & 2657.10 & 2162.93 \\
3 & 32 & 36 & 1152 & 87 & 2619.75 & 2270.94 \\
3 & 32 & 36 & 1152 & 521 & 7166.46 & 6520.03 \\
3 & 32 & 36 & 1152 & 869 & 10802.19 & 10563.85 \\
    \bottomrule
    \end{tabular}
    \end{center}
    \end{small}
\end{table}

\begin{table}
  \caption{Results for $k$-ported Alltoall for $k=4,5,6$ on the
    ``Hydra'' system.  The MPI library used is \intelmpiversion.}
    \label{tab:alltoall.port.n32.intel.B}
    \begin{small}
    \begin{center}
    \begin{tabular}{rrrrrrr}
      \toprule
$k$ & $n$ & $N$ & $p$ & $c$ & avg ($\mu s$) & min ($\mu s$) \\
      \midrule
      \multicolumn{7}{c}{Alltoall, $4$-ported} \\
4 & 32 & 36 & 1152 & 1 & 1874.32 & 1651.05 \\
4 & 32 & 36 & 1152 & 6 & 2210.83 & 1863.96 \\
4 & 32 & 36 & 1152 & 9 & 2244.16 & 1880.88 \\
4 & 32 & 36 & 1152 & 53 & 2144.99 & 1954.08 \\
4 & 32 & 36 & 1152 & 87 & 2369.90 & 2143.14 \\
4 & 32 & 36 & 1152 & 521 & 6580.91 & 6482.12 \\
4 & 32 & 36 & 1152 & 869 & 10745.88 & 10547.88 \\
      \midrule
      \multicolumn{7}{c}{Alltoall, $5$-ported} \\
5 & 32 & 36 & 1152 & 1 & 1896.75 & 1479.15 \\
5 & 32 & 36 & 1152 & 6 & 2084.02 & 1728.06 \\
5 & 32 & 36 & 1152 & 9 & 2137.49 & 1777.17 \\
5 & 32 & 36 & 1152 & 53 & 2163.66 & 1868.96 \\
5 & 32 & 36 & 1152 & 87 & 2295.80 & 2027.99 \\
5 & 32 & 36 & 1152 & 521 & 6749.93 & 6492.14 \\
5 & 32 & 36 & 1152 & 869 & 10995.11 & 10546.92 \\
      \midrule
      \multicolumn{7}{c}{Alltoall, $6$-ported} \\
6 & 32 & 36 & 1152 & 1 & 1704.25 & 1404.05 \\
6 & 32 & 36 & 1152 & 6 & 1914.40 & 1429.08 \\
6 & 32 & 36 & 1152 & 9 & 2016.79 & 1565.93 \\
6 & 32 & 36 & 1152 & 53 & 2019.58 & 1863.96 \\
6 & 32 & 36 & 1152 & 87 & 2169.93 & 2010.11 \\
6 & 32 & 36 & 1152 & 521 & 6748.24 & 6489.04 \\
6 & 32 & 36 & 1152 & 869 & 10820.21 & 10561.94 \\
    \bottomrule
    \end{tabular}
    \end{center}
    \end{small}
\end{table}

\begin{table}
  \caption{Results for full-lane Alltoall and the native \mpialltoall on the
    ``Hydra'' system.  The MPI library used is \intelmpiversion.}
    \label{tab:alltoall.full.n32.intel}
    \begin{small}
    \begin{center}
    \begin{tabular}{rrrrrrr}
      \toprule
$k$ & $n$ & $N$ & $p$ & $c$ & avg ($\mu s$) & min ($\mu s$) \\
      \midrule
      \multicolumn{7}{c}{Full-lane Alltoall} \\
6 & 32 & 36 & 1152 & 1 & 114.07 & 97.99 \\
6 & 32 & 36 & 1152 & 6 & 152.37 & 139.95 \\
6 & 32 & 36 & 1152 & 9 & 195.05 & 176.91 \\
6 & 32 & 36 & 1152 & 53 & 804.12 & 766.99 \\
6 & 32 & 36 & 1152 & 87 & 1356.17 & 1307.96 \\
6 & 32 & 36 & 1152 & 521 & 6690.57 & 6545.07 \\
6 & 32 & 36 & 1152 & 869 & 10700.44 & 10462.05 \\
      \midrule
      \multicolumn{7}{c}{\mpialltoall} \\
6 & 32 & 36 & 1152 & 1 & 214.64 & 200.03 \\
6 & 32 & 36 & 1152 & 6 & 389.15 & 351.91 \\
6 & 32 & 36 & 1152 & 9 & 423.43 & 403.88 \\
6 & 32 & 36 & 1152 & 53 & 1679.47 & 1633.88 \\
6 & 32 & 36 & 1152 & 87 & 4285.99 & 4120.83 \\
6 & 32 & 36 & 1152 & 521 & 8502.46 & 8337.02 \\
6 & 32 & 36 & 1152 & 869 & 12877.60 & 12515.07 \\
    \bottomrule
    \end{tabular}
    \end{center}
    \end{small}
\end{table}

The alltoall results with \mpichversion are shown in
Table~\ref{tab:alltoall.lane.n32.mpich},
Table~\ref{tab:alltoall.port.n32.mpich},
Table~\ref{tab:alltoall.port.n32.mpich.B} and
Table~\ref{tab:alltoall.full.n32.mpich}.  The results are similar to
the results with \openmpiversion and \intelmpiversion.

\begin{table}
  \caption{Results for $k$-lane Alltoall for $k=32$ on the
    ``Hydra'' system.  The MPI library used is \mpichversion.}
    \label{tab:alltoall.lane.n32.mpich}
    \begin{small}
    \begin{center}
    \begin{tabular}{rrrrrrr}
      \toprule
$k$ & $n$ & $N$ & $p$ & $c$ & avg ($\mu s$) & min ($\mu s$) \\
    \multicolumn{7}{c}{Alltoall, $32$ virtual lanes} \\
1 & 32 & 36 & 1152 & 1 & 983.41 & 866.17 \\
1 & 32 & 36 & 1152 & 6 & 1198.57 & 1116.28 \\
1 & 32 & 36 & 1152 & 9 & 1226.21 & 1110.08 \\
1 & 32 & 36 & 1152 & 53 & 1905.96 & 1810.07 \\
1 & 32 & 36 & 1152 & 87 & 2063.05 & 1973.15 \\
1 & 32 & 36 & 1152 & 521 & 7300.25 & 6990.19 \\
1 & 32 & 36 & 1152 & 869 & 11780.65 & 11185.65 \\
    \bottomrule
    \end{tabular}
    \end{center}
    \end{small}
\end{table}

\begin{table}
  \caption{Results for $k$-ported Alltoall for $k=1,2,3$ on the
    ``Hydra'' system.  The MPI library used is \mpichversion.}
    \label{tab:alltoall.port.n32.mpich}
    \begin{small}
    \begin{center}
    \begin{tabular}{rrrrrrr}
      \toprule
$k$ & $n$ & $N$ & $p$ & $c$ & avg ($\mu s$) & min ($\mu s$) \\
      \midrule
      \multicolumn{7}{c}{Alltoall, $1$-ported} \\
1 & 32 & 36 & 1152 & 1 & 3242.40 & 2975.23 \\
1 & 32 & 36 & 1152 & 6 & 4077.54 & 3657.58 \\
1 & 32 & 36 & 1152 & 9 & 4230.32 & 3809.93 \\
1 & 32 & 36 & 1152 & 53 & 5904.46 & 4013.06 \\
1 & 32 & 36 & 1152 & 87 & 4459.94 & 4273.65 \\
1 & 32 & 36 & 1152 & 521 & 7835.58 & 7612.23 \\
1 & 32 & 36 & 1152 & 869 & 11232.73 & 11008.26 \\
      \midrule
      \multicolumn{7}{c}{Alltoall, $2$-ported} \\
2 & 32 & 36 & 1152 & 1 & 2705.75 & 2305.27 \\
2 & 32 & 36 & 1152 & 6 & 3342.93 & 3069.40 \\
2 & 32 & 36 & 1152 & 9 & 3273.68 & 2998.59 \\
2 & 32 & 36 & 1152 & 53 & 3197.99 & 2729.18 \\
2 & 32 & 36 & 1152 & 87 & 3164.00 & 2902.75 \\
2 & 32 & 36 & 1152 & 521 & 8287.33 & 6599.19 \\
2 & 32 & 36 & 1152 & 869 & 10788.35 & 10582.45 \\
      \midrule
      \multicolumn{7}{c}{Alltoall, $3$-ported} \\
3 & 32 & 36 & 1152 & 1 & 2294.24 & 2058.74 \\
3 & 32 & 36 & 1152 & 6 & 2862.06 & 2625.94 \\
3 & 32 & 36 & 1152 & 9 & 2897.13 & 2688.65 \\
3 & 32 & 36 & 1152 & 53 & 2536.49 & 2298.59 \\
3 & 32 & 36 & 1152 & 87 & 2737.10 & 2436.64 \\
3 & 32 & 36 & 1152 & 521 & 6726.01 & 6550.55 \\
3 & 32 & 36 & 1152 & 869 & 10967.96 & 10607.00 \\
    \bottomrule
    \end{tabular}
    \end{center}
    \end{small}
\end{table}

\begin{table}
  \caption{Results for $k$-ported Alltoall for $k=4,5,6$ on the
    ``Hydra'' system.  The MPI library used is \mpichversion.}
    \label{tab:alltoall.port.n32.mpich.B}
    \begin{small}
    \begin{center}
    \begin{tabular}{rrrrrrr}
      \toprule
$k$ & $n$ & $N$ & $p$ & $c$ & avg ($\mu s$) & min ($\mu s$) \\
      \midrule
      \multicolumn{7}{c}{Alltoall, $4$-ported} \\
4 & 32 & 36 & 1152 & 1 & 2294.85 & 1941.20 \\
4 & 32 & 36 & 1152 & 6 & 2649.18 & 2386.09 \\
4 & 32 & 36 & 1152 & 9 & 2737.78 & 2414.94 \\
4 & 32 & 36 & 1152 & 53 & 2384.60 & 2114.06 \\
4 & 32 & 36 & 1152 & 87 & 2675.48 & 2296.45 \\
4 & 32 & 36 & 1152 & 521 & 7999.99 & 6505.97 \\
4 & 32 & 36 & 1152 & 869 & 10696.80 & 10539.05 \\
      \midrule
      \multicolumn{7}{c}{Alltoall, $5$-ported} \\
5 & 32 & 36 & 1152 & 1 & 1979.67 & 1740.93 \\
5 & 32 & 36 & 1152 & 6 & 2468.09 & 2304.08 \\
5 & 32 & 36 & 1152 & 9 & 2484.59 & 2199.65 \\
5 & 32 & 36 & 1152 & 53 & 2203.85 & 2032.76 \\
5 & 32 & 36 & 1152 & 87 & 2366.56 & 2176.76 \\
5 & 32 & 36 & 1152 & 521 & 6606.52 & 6499.77 \\
5 & 32 & 36 & 1152 & 869 & 10697.68 & 10550.26 \\
      \midrule
      \multicolumn{7}{c}{Alltoall, $6$-ported} \\
6 & 32 & 36 & 1152 & 1 & 1895.56 & 1664.40 \\
6 & 32 & 36 & 1152 & 6 & 2339.31 & 2098.08 \\
6 & 32 & 36 & 1152 & 9 & 2373.41 & 2155.07 \\
6 & 32 & 36 & 1152 & 53 & 2138.45 & 1975.30 \\
6 & 32 & 36 & 1152 & 87 & 2347.14 & 2154.11 \\
6 & 32 & 36 & 1152 & 521 & 6665.05 & 6489.04 \\
6 & 32 & 36 & 1152 & 869 & 10914.50 & 10557.89 \\
    \bottomrule
    \end{tabular}
    \end{center}
    \end{small}
\end{table}

\begin{table}
  \caption{Results for full-lane Alltoall and the native \mpialltoall on the
    ``Hydra'' system.  The MPI library used is \mpichversion.}
    \label{tab:alltoall.full.n32.mpich}
    \begin{small}
    \begin{center}
    \begin{tabular}{rrrrrrr}
      \toprule
$k$ & $n$ & $N$ & $p$ & $c$ & avg ($\mu s$) & min ($\mu s$) \\
      \midrule
      \multicolumn{7}{c}{Full-lane Alltoall} \\
6 & 32 & 36 & 1152 & 1 & 120.81 & 108.72 \\
6 & 32 & 36 & 1152 & 6 & 173.17 & 158.55 \\
6 & 32 & 36 & 1152 & 9 & 209.83 & 189.07 \\
6 & 32 & 36 & 1152 & 53 & 713.44 & 683.07 \\
6 & 32 & 36 & 1152 & 87 & 1376.79 & 1164.44 \\
6 & 32 & 36 & 1152 & 521 & 7953.45 & 6697.42 \\
6 & 32 & 36 & 1152 & 869 & 10851.71 & 10731.46 \\
      \midrule
      \multicolumn{7}{c}{\mpialltoall} \\
6 & 32 & 36 & 1152 & 1 & 456.90 & 434.88 \\
6 & 32 & 36 & 1152 & 6 & 570.17 & 559.81 \\
6 & 32 & 36 & 1152 & 9 & 675.07 & 660.18 \\
6 & 32 & 36 & 1152 & 53 & 1680.53 & 1642.94 \\
6 & 32 & 36 & 1152 & 87 & 2219.15 & 2002.48 \\
6 & 32 & 36 & 1152 & 521 & 6847.58 & 6660.70 \\
6 & 32 & 36 & 1152 & 869 & 11261.56 & 10731.46 \\
    \bottomrule
    \end{tabular}
    \end{center}
    \end{small}
\end{table}

\clearpage

\section{Conclusion}

This note discussed the $k$-lane communication model in which $k$
processors on a compute node can communicate simultaneously without
communication degradation, and raise the question of how to design
algorithms for standard, collective operations that can exploit these
capabilities. The $k$-lane model is different from the often used,
more powerful $k$-ported model, in which \emph{each processor} can at
a time be involved in up to $k$ communication operations. Different
approaches to designing good $k$-lane algorithms for broadcast,
scatter and alltoall were explored. Another, possible approach using
dynamic programming to construct optimal trees as in
\cite{Traff19:optimalgatherscatter} was not discussed here.
Unfortunately, the results so far are inconclusive, with rarely any
definite advantage over standard $k$-ported implementations, even for
$k=1$. The results, as is often the case, however, show many instances
where the native MPI library implementations of the collective
operations (\mpibcast, \mpiscatter, \mpialltoall) can easily be
improved, and sometimes quite considerably.  We think it
deserves to explore the $k$-lane model further.

\bibliographystyle{plain}
\bibliography{traff,parallel} 

\begin{thebibliography}{10}

\bibitem{Bar-Noy95}
Amotz Bar-Noy, Jehoshua Bruck, Ching-Tien Ho, Schlomo Kipnis, and Baruch
  Schieber.
\newblock Computing global combine operations in the multiport postal model.
\newblock {\em {IEEE} Transactions on Parallel and Distributed Systems},
  6(8):896--900, 1995.

\bibitem{BarNoyHo99}
Amotz Bar-Noy and Ching-Tien Ho.
\newblock Broadcasting multiple messages in the multiport model.
\newblock {\em {IEEE} Transactions on Parallel and Distributed Systems},
  10(5):500--508, 1999.

\bibitem{Bruck97}
Jehoshua Bruck, Ching-Tien Ho, Schlomo Kipnis, Eli Upfal, and D.~Weathersby.
\newblock Efficient algorithms for all-to-all communications in multiport
  message-passing systems.
\newblock {\em {IEEE} Transactions on Parallel and Distributed Systems},
  8(11):1143--1156, 1997.

\bibitem{ChanHeimlichPurkayasthavandeGeijn07}
Ernie Chan, Marcel Heimlich, Avi Purkayastha, and Robert~A. van~de Geijn.
\newblock Collective communication: theory, practice, and experience.
\newblock {\em Concurrency and Computation: Practice and Experience},
  19(13):1749--1783, 2007.

\bibitem{Jia09}
Bin Jia.
\newblock Process cooperation in multiple message broadcast.
\newblock {\em {P}arallel {C}omputing}, 35(12):572--580, 2009.

\bibitem{MPI-3.1}
{MPI Forum}.
\newblock {\em \textsf{MPI}: A Message-Passing Interface Standard. Version
  3.1}, June 4th 2015.
\newblock \url{www.mpi-forum.org}.

\bibitem{SackGropp15}
Paul Sack and William Gropp.
\newblock Collective algorithms for multiported torus networks.
\newblock {\em {ACM} Transactions on Parallel Computing}, 1(2):12:1--12:33,
  2015.

\bibitem{Traff19:lanecorr}
Jesper~Larsson Tr{\"a}ff.
\newblock Decomposing collectives for exploiting multi-lane communication.
\newblock arXiv:1910.13373, 2019.

\bibitem{Traff19:optimalgatherscatter}
Jesper~Larsson Tr{\"a}ff.
\newblock On optimal trees for irregular gather and scatter collectives.
\newblock {\em {IEEE} Transactions on Parallel and Distributed Systems},
  30(9):2060--2074, 2019.

\bibitem{Traff20:mpidecomp}
Jesper~Larsson Tr{\"a}ff and Sascha Hunold.
\newblock Decomposing {MPI} collectives for exploiting multi-lane
  communication.
\newblock In {\em {IEEE} {CLUSTER}}. IEEE Computer Society, 2020.

\bibitem{Traff08:optibcast}
Jesper~Larsson Tr{\"a}ff and Andreas Ripke.
\newblock Optimal broadcast for fully connected processor-node networks.
\newblock {\em Journal of Parallel and Distributed Computing}, 68(7):887--901,
  2008.

\bibitem{Traff14:bruck}
Jesper~Larsson Tr{\"a}ff, Antoine Rougier, and Sascha Hunold.
\newblock Implementing a classic: Zero-copy all-to-all communication with {MPI}
  datatypes.
\newblock In {\em 28th {ACM} International Conference on Supercomputing
  ({ICS})}, pages 135--144. ACM, 2014.

\end{thebibliography}

\end{document}